\documentclass[%
preprint,
amsmath,amssymb,
aps,
prb,
floatfix,
superscriptaddress,
]{revtex4-2}
\usepackage{graphicx}
\usepackage{dcolumn}
\usepackage{bm}
\usepackage{hyperref}

\usepackage{multirow}
\usepackage{graphicx}
\usepackage{xcolor}

\newcommand{\mb}{\mathbf}
\newcommand{\mbh}[1]{\hat{\mathbf{#1}}}
\newcommand{\bs}{\boldsymbol} 
\newcommand{\mbr}{\mathbf{r}}
\newcommand{\bsrho}{\boldsymbol{\rho}}
\newcommand{\T}{\text}

\newcommand{\mc}[1]{\mathcal{#1}}

\newcommand{\alphabar}{\bar{\alpha}}

\begin{document}

\title{Theory for tip-enhanced Raman spectroscopy of two-dimensional materials including out-of-plane Raman response}

\author{Raul Corr\^ea}\email{raulcs@fisica.ufmg.br}
\altaffiliation[Present address: ]{Instituto de F\'isica, Universidade Federal do Rio de Janeiro, Rio de Janeiro, RJ 21941-972, Brazil}
\affiliation{Departamento de F\'isica, Universidade Federal de Minas Gerais, Belo Horizonte, MG 30123-970, Brazil}
\author{Luiz G. Can\c cado}
\author{Ado Jorio}\email{adojorio@fisica.ufmg.br}
\affiliation{Departamento de F\'isica, Universidade Federal de Minas Gerais, Belo Horizonte, MG 30123-970, Brazil}

\date{\today}

\begin{abstract}
    Tip-Enhanced Raman Spectroscopy (TERS) can be used to make nanoscale spatial measurements of 2D materials, such as graphene and transition metal dichalcogenides (TMDs).
    The TERS theory introduced in [L. Can\c cado 
    \textit{et al.}, \href{https://doi.org/10.1103/PhysRevX.4.031054}{Phys. Rev. X \textbf{4}, 031054 (2014)}], however, was tailored for graphene, whose out-of-plane Raman response is neglected.
    In the present work, we include the out-of-plane response in the TERS theory.
    In doing so, we provide an exact analytical expression for the field propagation between the tip and the sample,
    and show that
    the contribution to the TERS signal that scatters first at the sample, then at the tip (sample-tip, or TS)
    is important only when the out-of-plane response is significant.
    We extensively study the variation of TERS experimental measurements when varying physical parameters of the system, like the tip radius, the out-of-plane response, the TERS coherence length, and others.
    It becomes evident that the TERS enhancement is very sensitive to the out-of-plane Raman response of the phonon mode,
    while normalized tip-approach measurements are more sensitive to the coherence length,
    and we show that the medium refractive index leads to an effective tip enhancement factor $f_e$.
    Our results lead to the conclusion that, in general, a strong TERS enhancement is a necessary condition for investigating the physics discussed here,
    which here means surveying the difference in TERS signals between different Raman modes.
    We use our model to analyze some graphene TERS experiments, showing that they are consistent with a negligible out-of-plane Raman response and a nonzero TERS coherence length in the fitting.
\end{abstract}

\maketitle

\section{Introduction}

Raman spectroscopy is a powerful well-established technique to investigate physical and chemical properties of materials,
in which light is scattered inelastically by interaction with material degrees of freedom, to frequencies adjacent to the excitation field~\cite{cardona1,bloembergen}.
When exciting with visible light, these frequency changes are very small, so resolution-wise one is optically limited by the excitation wavelength, that is tenths of micrometers.
At this resolution, small spatial coherences in the electric polarization of the material, on the order of nanometers, can be safely neglected, and Raman scattering can be treated as a spatially incoherent process.
There are, however, techniques to improve that resolution, and Tip-Enhanced Raman Spectroscopy (TERS) is one that can reach the nanometers scale~\cite{novotny,kawata_shalaev,jorio2024,bao2024}.
In this method, a sharp metallic object is placed close to the measured sample, so close that the strong near field that the tip emits in response to the laser greatly enhances the Raman signal scattered by the sample.
As the enhancement only happens in the nanometric region around the tip, one thus obtains both a nanometrically resolved and a greatly enhanced Raman spectroscopic signal.

In the TERS regime, phenomena at the nanometric scale start to play a role in the scattering, in particular if they present spatial coherence.
In that case, the symmetry of the excited phonon modes along a wide region of the material can lead to constructive or destructive interferences that affect the TERS signal enhancement in relation to regular far field Raman scattering, in which atoms respond independently of each other.
This phenomenon has been experimentally observed in graphene~\cite{beams2014, nadas2025} and GaS~\cite{alencar2019}, and a theory has been developed for it assuming that its Raman response in the out-of-plane axis
(that is, in the direction perpendicular to the material plane),
characterized by the corresponding Raman tensor component, is null~\cite{cancado2014,publio2022}.
It is known that this component is not null for transition metal dichalcogenides (TMDs) \cite{raman_tensor_mose2}, because these materials are not as flat along it as is graphene,
so the TERS theory for 2D materials must be extended in order to describe TERS measurements on TMDs \cite{jorio2024}.
On top of that, one must consider the possibility that the near field of the tip, which has a much stronger spatial gradient than the far field of the excitation laser, might even lead to an out-of-plane response in graphene \cite{pratama2019}.

In this article, we extend the previous TERS theory for probing 2D materials \cite{cancado2014,publio2022} in three important aspects:
(i) the inclusion of the out-of-plane Raman response in the TERS signal, which greatly modifies the contributions to be considered;
(ii) we provide the exact analytical solution of the tip-sample-tip (TST), tip-sample (ST) and sample-tip (TS) field propagators, which have previously been obtained with an approximation \cite{cancado2014,publio2022}, including the out-of-plane contribution;
(iii) we show that the TS contribution may be neglected in comparison with the TST and ST terms, when the excitation field is a focused radially polarized Gaussian beam, but only if the out-of-plane sample response is null.
In addition, a discussion on the influence of the physical parameters of the model in the TERS intensity signal is made, in a more extensive manner than in previous works,
including an analysis of how the medium refractive index leads to an effective tip enhancement factor $f_e$.
We will argue that, in general, a strong enhancement is needed to distinguish some physical property affecting the behavior of Raman mode intensities in TERS,
be it a large $f_e$, tip radius, relative Raman response or placing the tip very close to the sample.
We also use our extended theory to show that graphene TERS experiments seem to strengthen the idea that its out-of-plane Raman response is indeed negligible.

The article is organized as follows.
In Sec.~\ref{sec:ters_scattering}, we introduce the scattering framework used in our TERS theory,
while in Sec.~\ref{sec:polarizability} the model for the spatial correlation of the material electric polarizability due to the phonon excitations, which defines the coherence length measured by TERS, is discussed.
Section~\ref{sec:LcJ} presents an interpretation of the TERS scattering as a propagation spatial frequency filter on the polarizability correlation.
Section~\ref{sec:propagation} develops the analytical expression of the propagators of TST, ST and TS components of TERS,
and in Sec.~\ref{sec:ters_of_modes} these are applied to different phonon modes, where the influence of the out-of-plane response is presented.
Section~\ref{sec:discussion} provides an extensive discussion of our results, with an in-depth analysis of the change in TERS measurements by the physical parameters of the system, given our model,
a comparison with graphene experimental data,
and a discussion on spatial correlation functions other than the one we use in our model.
We conclude in Sec.~\ref{sec:conclusion}.

\section{TERS intensity in a scattering framework}\label{sec:ters_scattering}

\begin{figure}[!ht]
\begin{center}
    \includegraphics[width=.6\textwidth]{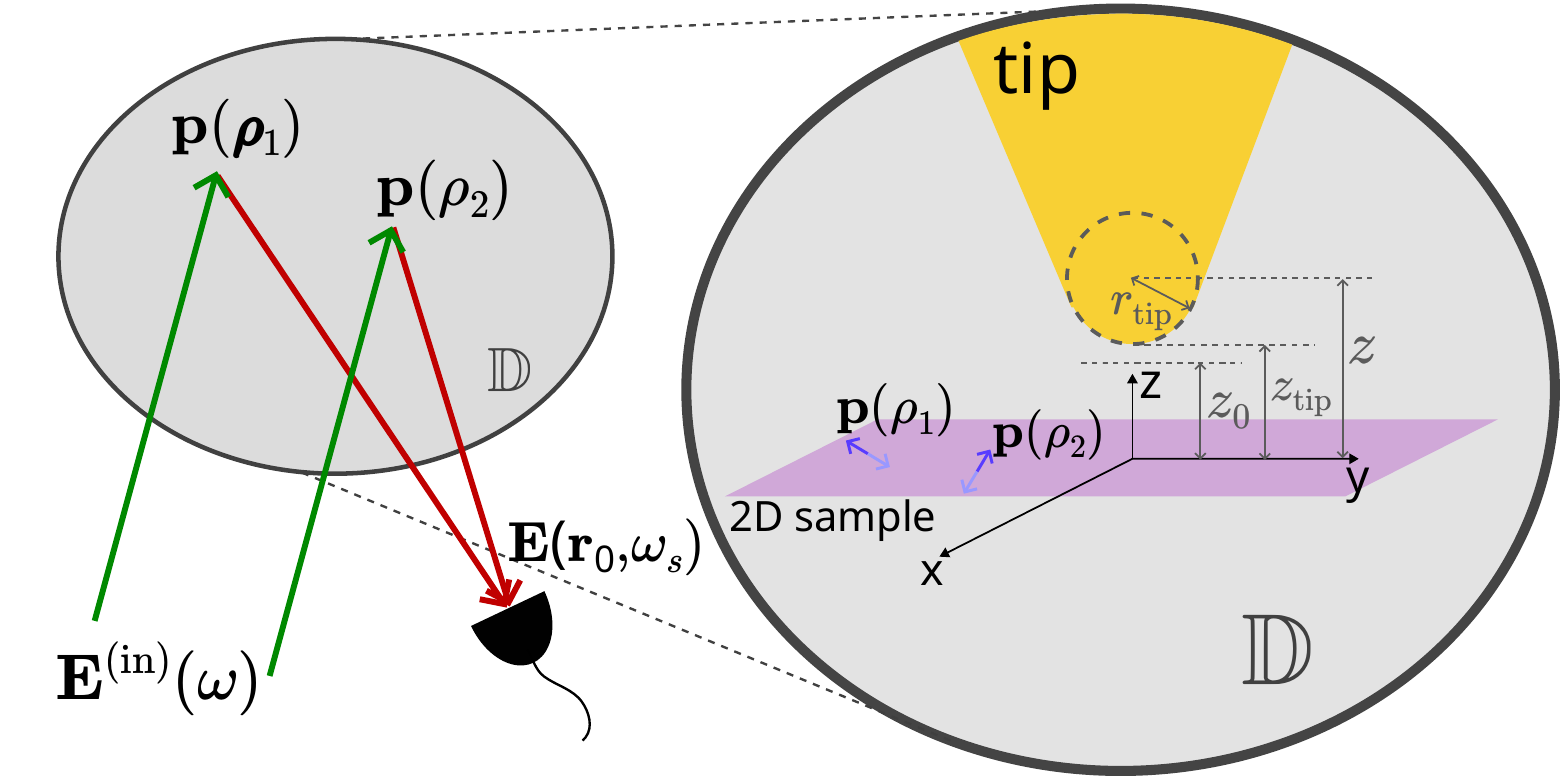}
\end{center}
    \caption{\label{fig:scattering}General scattering framework used in our TERS theory.
    A monochromatic excitation field
    $\mb{E}^{\T{(in)}}(\omega)$, oscillating at frequency $\omega$,
    reaches a scattering medium with domain $\mathbb{D}$,
    and is scattered as the field $\mb{E}(\mb{r}_0,\omega_s)$ to a detector in $\mbr_0$ and at a frequency $\omega_s$.
    To get the scattered field intensity, we coherently sum the contribution of each pairs of polarizations $\mb{p}(\bs{\rho}_1)$ and $\mb{p}(\bs{\rho}_2)$, excited at points $\bs{\rho}_1$ and $\bs{\rho}_2$ in $\mathbb{D}$.
    The scattering medium $\mathbb{D}$ contains both the sample and the tip, as well as the optics to collect the scattered field.
    In the zoom circle, we represent two dipoles $\mb{p}(\bsrho_1)$ and $\mb{p}(\bsrho_2)$ in the sample with nonzero in-plane and out-of-plane components, as a response to the incident and tip electric fields.
    We define $z_{\T{tip}}$ as the distance from the sample plane to the point in the tip closest to the sample,
    and $z_0$ as the smallest $z_{\T{tip}}$ in a TERS experiment.
    The tip-sample distance $z$ appearing in the theory localizes the tip emitting dipole inside the tip at a distance $r_{\T{tip}}$ from the tip surface,
    where $r_{\T{tip}}$ is the tip radius,
    so $z=z_{\T{tip}}+r_{\T{tip}}$.
    }
\end{figure}

Tip-Enhanced Raman Spectroscopy (TERS) of 2D materials can be modelled by the scattering scheme of Fig.~\ref{fig:scattering}.
Our treatment will resemble that of Ref.~\cite{cancado2014}, so that we can keep a comparison.
A monochromatic electric field $E_i^{(\T{in})}(\mbr,\omega)$ ($i\in\{x,y,z\}$ is an index for a component of the field) with frequency $\omega$ reaches the scattering system composed of the material sample and the tip, and is scattered from there towards a detector far from it.
The sample defines the $xy$ plane, and the tip axis, which is perpendicular to the sample plane in our geometry, defines the $z$ axis.
In the notation of Fig.~\ref{fig:scattering}, $z_0$ is the closest distance that the tip can be to the sample, when we say that the tip is engaged;
$z_{\T{tip}}$ is some tip-sample distance larger than $z_0$;
$r_{\T{tip}}$ is the tip radius;
and $z=z_{\T{tip}}+r_{\T{tip}}$ defines the distance on the tip axis from the sample to an emitting dipole in the center of the tip, used to model the field scattered by it.

Due to the particular configuration of this two-part system, a series of scattering events can happen.
Namely, the field can be scattered by just the sample and go to the detector (this term is represented by S);
or by just the tip (term T);
or be scattered by the tip, then the sample, then go to the detector, which is the tip-sample term (ST in the usual operator notation, right happens first);
or first in the sample and then in the tip, the sample-tip term (TS);
or it can be a tip-sample-tip event (TST),
and so on.
This is represented by the series \cite{novotny}
\begin{equation*}
    \T{T} + \T{S} + \T{ST} + \T{TS} + \T{TST} + \dots
\end{equation*}
The T term does not probe the sample, and S is just usual Raman without the tip.
The TERS is present in terms ST, TS, and TST, and higher-order terms in the series can be neglected for being too weak at the distances at which the tip can approach the sample.

A monochromatic field with frequency $\omega$ excites the sample at point $\bsrho$ (we use the Greek letter $\bsrho$ to make it explicit that we are in the 2D plane of the sample), whose $i$th component is represented by $E_i(\bsrho,\omega)$.
It may have come from the tip, which is the case for ST and TST terms,
or it may have come straight from the source [$E_i^{(\T{in})}(\mbr,\omega)$], as in the TS term.
It excites an electric polarization $p^{(\eta)}_j (\bsrho,\omega_s;\omega)$ at that point $\bsrho$ with frequency $\omega_s$, in response to $\omega$, due to some internal (vibrational) mode $\eta$,
whose response is given by a polarizability tensor $\alpha^{(\eta)}_{ij}$
\footnote{We do not use the repeated index summation convention because it may get ambiguous later.},
\begin{equation}
    p^{(\eta)}_j (\bsrho,\omega_s; \omega)
    =
    \sum_i \alpha^{(\eta)}_{ij} (\bsrho; \omega_s, \omega)
    E_i (\bsrho,\omega).
\end{equation}
This polarization may then create a field propagating straight to the detector (ST), located at point $\mbr_0$, or it may be scattered by the tip before it goes (TS and TST).
This creates the field $E^{(\eta)}_k (\mbr_0,\omega_s)$ at $\mbr_0$ with frequency $\omega_s$.
We write
\begin{equation}
    E^{(\eta)}_k (\mbr_0,\omega_s)
    =
    \frac{\omega_s^2}{\varepsilon_0 c}
    \sum_j
    \int_\mathbb{D}
        G_{kj}(\mbr_0,\bsrho;\omega_s)
        p^{(\eta)}_j (\bsrho,\omega_s; \omega)
    d^2\bsrho,
\end{equation}
where $G_{kj}(\mbr_0,\bsrho;\omega_s)$ is a dyadic Green function containing all the eventual intermediate scatterings between sample and detector.

Since the stationary polarization response can be taken as instantaneous,
$p^{(\eta)}_j (\bsrho,t) = \alpha^{(\eta)}_{ij} (\bsrho,t) E_i (\bsrho,t)$,
it implies
$\alpha^{(\eta)}_{ij} (\bsrho; \omega_s, \omega)
= \alpha^{(\eta)}_{ij} (\bsrho, \omega-\omega_s)$.
Then the scattered field component $k$ due to internal material mode $\eta$ is
\begin{eqnarray}
    E^{(\eta)}_k (\mbr_0,\omega_s)
    =&&
    \frac{\omega_s^2}{\varepsilon_0 c}
    \sum_{i,j}
    \int_\mathbb{D}
        G_{kj}(\mbr_0,\bsrho;\omega_s)
    \notag
   \\&&\times
        \alpha^{(\eta)}_{ij} (\bsrho, \omega-\omega_s) E_i (\bsrho, \omega)
    d^2 \bsrho.
\end{eqnarray}

The scattered field is stationary, so its Fourier transform $E_k(\mbr_0,\omega_s)$ is ill-defined,
but its spectral density $S(\mbr_0,\omega_s)$ is a well-behaved physical quantity, if defined according to
$S(\mbr_0,\omega_s)\delta(\omega_s'-\omega_s) \equiv \sum_k \langle E_k^*(\mbr_0,\omega_s) E_k(\mbr_0,\omega_s')\rangle$
\cite{mandelwolf}.
Then,
we can define
\begin{equation}\label{eq:F_definition}
    F_{ijk} (\mb{r}; \mbr_0, \omega_s,\omega)
    \equiv
    G_{kj}(\mbr_0,\bsrho;\omega_s) E_i (\bsrho, \omega),
\end{equation}
such that the
scattered spectral density
at point $\mb{r}_0$ and frequency $\omega_s$ is
\begin{equation}\label{eq:S_FA}
    S^{(\eta)} (\mbr_0, \omega_s)
    = \frac{\omega_s^4}{\varepsilon_0^2 c^2}
    \sum_{i,i',j,j',k}
    \iint_\mathbb{D}
    F_{ijk}^* (\bsrho_1; \mbr_0, \omega_s,\omega)
    F_{i'j'k} (\bsrho_2; \mbr_0, \omega_s,\omega)
    A^{(\eta)}_{ii',jj'} (\bsrho_2-\bsrho_1, \omega-\omega_s)
    d^2\bsrho_1  d^2\bsrho_2
    ,
\end{equation}
where we used
\begin{equation}\label{eq:polarizability_correlation}
    A^{(\eta)}_{ii',jj'} (\bsrho_2-\bsrho_1, \omega-\omega_s)
    \delta(\omega_s'-\omega_s)
    \equiv
    \langle
        \alpha^{(\eta)*}_{ij} (\bsrho_1, \omega-\omega_s)
        \alpha^{(\eta)}_{i'j'} (\bsrho_2, \omega-\omega_s')
    \rangle
\end{equation}
for the two-point correlation function of the electric polarizability of the sampled material.
The latter is the crucial physical quantity when we are concerned with spatial coherence, because it carries the response correlation between different points in the sample, telling us which pairs of points respond together to the excitation, and which respond independently from each other.
We will assume, in particular, that the polarizability correlation depends on the absolute difference $|\bsrho_2-\bsrho_1|$, being an isotropic quantity.

By writing the TERS scattered signal in the form of Eq.~(\ref{eq:S_FA}), making use of the fact that we have at most one sample scattering event,
the $F$ functions carry all the propagation information (including the tip-sample geometry from the terms ST, TS and TST) plus the incident electric field, but no phonon information.
That is, $F$ contains properties which should be constant across experiments with the same setup but different sampled materials.
All the information containing the sample response, like the symmetry of the probed phonon mode, is contained in the correlation function $A$.

In order to obtain analytical expressions, it will be more convenient to work in the Fourier space with domain $\tilde{\mathbb{D}}$, so we define the pairs
\begin{subequations}
\begin{equation}
    \tilde{F}_{ijk} (\mb{q}; \mbr_0, \omega_s,\omega)
    =
    \int_\mathbb{D}
        F_{ijk} (\bsrho; \mbr_0, \omega_s,\omega)
    e^{-i \mb{q}\cdot \bsrho }
    d^2 \bsrho,
\end{equation}
\begin{equation}
    F_{ijk} (\bsrho; \mbr_0, \omega_s,\omega)
    =
    \int_{\tilde{\mathbb{D}}}
        \tilde{F}_{ijk} (\mb{q}; \mbr_0, \omega_s,\omega)
    e^{i \mb{q}\cdot \bsrho }
    \frac{d^2 \mb{q}}{(2\pi)^2},
\end{equation}
\end{subequations}
and
\begin{subequations}
\begin{eqnarray}
    \tilde{A}^{(\eta)}_{ii',jj'} (\mb{q}, \omega-\omega_s)
    =&&
    \int_\mathbb{D}
        A^{(\eta)}_{ii',jj'} (\bsrho_2-\bsrho_1, \omega-\omega_s)
    \notag\\&&\times
    e^{-i \mb{q}\cdot (\bsrho_2-\bsrho_1) }
    d^2 (\bsrho_2-\bsrho_1),
\end{eqnarray}
\begin{eqnarray}
    A^{(\eta)}_{ii',jj'} (\bsrho_2-\bsrho_1, \omega-\omega_s)
    =&&
    \int_{\tilde{\mathbb{D}}}
        \tilde{A}^{(\eta)}_{ii',jj'} (\mb{q}, \omega-\omega_s)
    \notag\\&&\times
    e^{i \mb{q}\cdot (\bsrho_2-\bsrho_1) }
    \frac{d^2\mb{q}}{(2\pi)^2}.
\end{eqnarray}
\end{subequations}
The TERS signal of Eq.~(\ref{eq:S_FA}) can then be written as
\begin{equation}\label{eq:detected_signal}
    S^{(\eta)} (\mbr_0, \omega_s)
    =
    \frac{\omega_s^4}{\varepsilon_0^2 c^2}
    \sum_{i,i',j,j',k}
    \int_{\tilde{\mathbb{D}}}
        \tilde{F}_{i'j'k}^* (\mb{q}; \mbr_0, \omega_s,\omega)
        \tilde{F}_{ijk} (\mb{q}; \mbr_0, \omega_s,\omega)
        \tilde{A}^{(\eta)}_{ii',jj'} (\mb{q}, \omega-\omega_s)
    \frac{d^2{\mb{q}}}{(2\pi)^2},
\end{equation}
and the TERS intensity detected in the laboratory should be proportional to $S^{(\eta)} (\mbr_0, \omega_s)$.

\section{Model for the polarizability correlation}\label{sec:polarizability}

The model for the polarizability correlation is at the heart of the main extension that we make in the present work in relation to Ref. \cite{cancado2014}, which is the inclusion of the out-of-plane polarizability response.
If we want to model $A^{(\eta)}_{ii',jj'} (\bsrho_2-\bsrho_1, \omega-\omega_s)$,
we need the polarizability $\alpha_{ij}^{(\eta)} (\bsrho,t)$ due to material vibrations [see Eq.~(\ref{eq:polarizability_correlation})], which is modelled with a first-order expansion of its dependence on the material nuclei positions $Q_k^{(\eta)} (\bsrho,t)$,
\begin{equation}\label{eq:alpha_ijk_def}
    \alpha_{ij}^{(\eta)} (\bsrho,t) = \bar{\alpha}_{ij}^{(\eta)}
    +
    \sum_{k}
    \alpha_{ijk}^{(\eta)}
    Q_k^{(\eta)} (\bsrho,t),
\end{equation}
where $\alpha_{ijk}^{(\eta)} \equiv (\partial\alpha_{ij}^{(\eta)} / \partial Q^{(\eta)}_k)|_{Q_k=0}$ is a third-rank Raman tensor of mode $\eta$.
We put an $\eta$ label on $Q_k$ because its dynamics depends on the excited mode.
Ignoring the constant part, we get
\begin{eqnarray}
    &&\langle
        \alpha_{ij}^{(\eta) *} (\bsrho_1,t_1)
        \alpha_{i'j'}^{(\eta)} (\bsrho_2,t_2)
    \rangle
    \notag\\
    &&=
    \sum_{k,k'}
    \alpha_{ijk}^{(\eta) *}
    \alpha_{i'j'k'}^{(\eta)}
    \langle
        Q_{k}^{(\eta)} (\bsrho_1,t_1)
        Q_{k'}^{(\eta)} (\bsrho_2,t_2)
    \rangle,
\end{eqnarray}
which can be very cumbersome to calculate from first principles, and we would need to know the exact dynamics of the driven phonons if we were to do it.

Instead, we adopt the phenomenological approach of Ref. \cite{cancado2014} and write, in the frequency domain,
\begin{eqnarray}\label{eq:A_def}
    &&A^{(\eta)}_{ii',jj'} (\bsrho_2-\bsrho_1, \omega-\omega_s)
    \notag\\
    &&=
    \alpha_{ij}^{(\eta) *}
    \alpha_{i'j'}^{(\eta)}
    \Omega^{(\eta)}(\omega-\omega_s)
    \mc{L}^{(\eta)}(|\bsrho_2-\bsrho_1|; L_c),
\end{eqnarray}
where the $\alpha_{ij}^{(\eta)}$ are the Raman tensors of mode $\eta$ that carry the tensorial part of the polarizability correlation,
$\Omega^{(\eta)}(\omega-\omega_s)$ carries the central frequency and spectral width of phonon mode $\eta$,
and $\mc{L}^{(\eta)}(|\bsrho_2-\bsrho_1|; L_c)$ is the spatial part, characterized by a single length parameter $L_c$.
The meaning of $L_c$ is the $|\bsrho_2-\bsrho_1|$ distance over which $\mc{L}$ is above a certain value, so it plays the part of the coherence length of the phononic electric polarization induced in the material,
and we call it the TERS coherence length.
As $L_c \rightarrow 0$, we expect that $\mc{L} \rightarrow \delta^{(2)}(|\bsrho_2-\bsrho_1|)$, and as $L_c \rightarrow \infty$, we expect that $\mc{L} \rightarrow C$, $C$ being a constant.
This is reasonable choice for a correlation function, because this limit can provide the totally uncorrelated scattering \cite{cancado2014} when plugging $\mc{L} \rightarrow \delta^{(2)}(|\bsrho_2-\bsrho_1|)$ into Eq.~(\ref{eq:S_FA}).

A possible choice for a phenomenological $\mc{L}$ fulfilling these properties is a Gaussian function with unit integral \cite{cancado2014},
\begin{subequations}\label{eq:Lc}
    \begin{equation}
        \mc{L}^{(\eta)}(|\bsrho_2-\bsrho_1|; L_c)
        =
        \frac{1}{\pi L_c^2} e^{-|\bsrho_2-\bsrho_1|^2/L_c^2},
    \end{equation}
    \begin{equation}
        \tilde{\mc{L}}^{(\eta)}(\mb{q}; L_c)
        =
    e^{-|\mb{q}|^2 L_c^2/4}.
    \end{equation}
\end{subequations}
This choice has the particularly important property that its norm, equal to $1/(\sqrt{2\pi} L_c)$, decreases as $L_c$ increases.
In Sec.~\ref{sec:LcJ} we will see that the TERS signal can be interpreted as the inner product between a function $\mc{J}^{(\eta)}$
and $\mc{L}^{(\eta)}$.
Then, for a fixed $\mc{J}^{(\eta)}$, which means all experimental parameters are fixed, as $L_c$ gets larger its inner product with $\mc{L}^{(\eta)}$ can only get smaller, because of the decreasing norm of the latter.
This implies that the TERS signal decreases as $L_c$ increases if we use Eq.~(\ref{eq:A_def}) and Eq.~(\ref{eq:Lc}) as the model for the polarizability correlation.
In Sec.~\ref{sec:correlation_functions} we show that this behavior is common to all spatial correlation functions that approach a delta function when $L_c\rightarrow 0$.

Phonons of modes A$_1$ and E$_{2g}$ can yield a strong Raman signal in both graphene and TMDs~\cite{tan_raman}, and can also show a significant enhancement in TERS~\cite{beams2014,park2016}.
Mode A$_1$ is viewed as a scalar mode in graphene~\cite{jorio_graphene}, but it comes from an out-of-plane oscillation in TMDs~\cite{zhao2013}.
From symmetry considerations, their Raman polarizability tensors can be written in the general forms
\begin{subequations}\label{eq:alpha_matrices}
    \begin{equation}
        \alpha_{ij}^{(A_1)}
        =
        \alphabar^{(A_1)}
        \begin{bmatrix}
            1 & 0 & 0 \\ 0 & 1 & 0 \\ 0 & 0 & \nu
        \end{bmatrix}_{ij}
        ,
    \end{equation}
    \begin{equation}
        \alpha^{(E_{2g}, x)}_{ij} = \alphabar^{(E_{2g})}
            \begin{bmatrix}
                0 & 1 & 0 \\ 1 & 0 & 0 \\ 0 & 0 & 0
            \end{bmatrix}_{ij}
        ,
    \end{equation}
    \vspace{5 pt}
    \begin{equation}
        \alpha^{(E_{2g}, y)}_{ij} = \alphabar^{(E_{2g})}
        \begin{bmatrix}
            1 & 0 & 0 \\ 0 & -1 & 0 \\ 0 & 0 & 0
        \end{bmatrix}_{ij}
        .
    \end{equation}
\end{subequations}
In comparison with Ref. \cite{cancado2014}, we included the possibility of a nonzero $z$ component of the polarizability in the A$_1$ Raman tensor as the $\nu$ parameter, which will demand the calculation of new terms in the TERS signal.
It was known that differences in the symmetry between the two modes lead to a different decay in the TERS signal between them, when varying the tip-sample distance $z$ \cite{beams2014,cancado2014,nadas2025},
and $\nu > 0$ will be yet another factor to influence these differences, which we will investigate.

For TMDs, $\nu > 0$ must certainly be included in the treatment, with a striking experimental value of 4 being found in the literature \cite{raman_tensor_mose2}.
In graphene, in regular Raman without the tip, the excitation field can be taken as constant in a region of the order of its wavelength,
and due to a $z$-axis symmetry of the charge distribution in the material, the $z$ contribution to the polarization excited by this constant field cancels out, yielding $\nu=0$.
However, in the presence of the near field emitted by the tip, which varies on a nanometer scale, this symmetry can be broken, because the charge displaced in the material at $z>0$ will be considerably different than that at $z<0$ \cite{pratama2019}.
This opens the possibility for $\nu > 0$ to be considered in graphene in a TERS setting, and we aim to investigate how this parameter may influence graphene TERS measurements.

\section{Propagation as a spatial frequency filter on the correlation}\label{sec:LcJ}

It is convenient to define a quantity $\mc{I}^{(\eta)}$ that contains the propagation and tensorial parts of the TERS signal of mode $\eta$ by
\begin{equation}\label{eq:mcI}
    \tilde{\mc{I}}^{(\eta)} (\mb{q}; \mb{r}_0, \omega_s, \omega) \equiv
        \sum_k
        \left( \sum_{i',j'} \tilde{F}_{i'j'k}^*(\mb{q}; \mb{r}_0, \omega_s, \omega) \alpha_{i'j'}^{(\eta) *} \right)
        \left( \sum_{i,j} \tilde{F}_{ijk}(\mb{q}; \mb{r}_0, \omega_s, \omega) \alpha_{ij}^{(\eta)} \right)
    .
\end{equation}
Then, with Eqs.~(\ref{eq:detected_signal}) and (\ref{eq:A_def}), the TERS signal is written as
\begin{equation}\label{eq:detected_signal_mcI}
    S^{(\eta)} (\mbr_0, \omega_s)
    =
    \frac{\omega_s^4}{\varepsilon_0^2 c^2}
    \Omega^{(\eta)}(\omega-\omega_s)
    \int_\mathbb{D}
        \tilde{\mc{I}}^{(\eta)} (\mb{q}; \mb{r}_0, \omega_s, \omega)
        \tilde{\mc{L}}^{(\eta)} (\mb{q}; L_c)
    \frac{d^2{\mb{q}}}{(2\pi)^2},
\end{equation}
with the dependence on the coherence length $L_c$ being isolated in $\mc{L}^{(\eta)}$.
This has the nice interpretation that the propagation-tensorial function $\tilde{\mc{I}}^{(\eta)} (\mb{q}; z)$ acts like a spatial frequency filter over the correlation $\tilde{\mc{L}}^{(\eta)} (\mb{q}; L_c)$, which is then integrated over all wave vectors $\mb{q}$.
In other words, the TERS signal is proportional to the inner product between propagation (which depends on $z$) and the polarizability correlation.

Taking one step further, we can use the fact that the isotropic $\mc{L}^{(\eta)}$ is a function of the modulus $|\mb{q}| \equiv q$, but not its angle $\varphi_q$, and defining
$\tilde{\mc{J}}^{(\eta)}(q;\mbr_0,\omega_s,\omega) \equiv \int_0^{2\pi} \tilde{\mc{I}}^{(\eta)}(\mb{q};\mbr_0,\omega_s,\omega) d\varphi_q$ yields
\begin{equation}\label{eq:detected_signal_mcJ}
    S^{(\eta)} (\mbr_0, \omega_s)
    =
    \frac{\omega_s^4}{\varepsilon_0^2 c^2}
    \frac{\Omega^{(\eta)}(\omega-\omega_s)}{(2\pi)^2}
    \int_0^\infty
        \tilde{\mc{J}}^{(\eta)} (q; z, \mb{r}_0, \omega_s, \omega)
        \tilde{\mc{L}}^{(\eta)} (q; L_c)
    q dq,
\end{equation}
where we explicitly state that $\tilde{\mc{J}}^{(\eta)}$ depends on the tip-sample distance $z$.
Then, from Parseval's theorem on cylindrically symmetric functions \cite{piessens_hankel}, the inner product can be taken with respect to the spatial variable $\rho \equiv |\bs{\rho}_2-\bs{\rho}_1|$,
\begin{equation}\label{eq:detected_signal_rho}
    S^{(\eta)} (\mbr_0, \omega_s)
    =
    \frac{\omega_s^4}{\varepsilon_0^2 c^2}
    \Omega^{(\eta)}(\omega-\omega_s)
    \int_0^\infty
        \mc{J}^{(\eta)} (\rho; z, \mb{r}_0, \omega_s, \omega)
        \mc{L}^{(\eta)} (\rho; L_c)
    \rho d\rho,
\end{equation}
where $\mc{J}^{(\eta)} (\rho; z, \mb{r}_0, \omega_s, \omega)$ is the zeroth-order inverse Hankel transform of $\tilde{\mc{J}}^{(\eta)} (q; z, \mb{r}_0, \omega_s, \omega)$,
\begin{equation}\label{eq:J_hankel}
    \mc{J}^{(\eta)} (\rho; z, \mb{r}_0, \omega_s, \omega)
    \equiv
    \frac{1}{2\pi}
    \int_0^\infty
        \tilde{\mc{J}}^{(\eta)} (q; z, \mb{r}_0, \omega_s, \omega)
        J_0(\rho q)
    q dq
    ,
\end{equation}
and analogously for $\mc{L}^{(\eta)} (\rho; L_c)$, whose expression is that in Eq.~(\ref{eq:Lc}) for $\rho = |\bs{\rho}_2-\bs{\rho}_1|$.
Note that, with the definition of Eq.~(\ref{eq:J_hankel}), the zeroth-order Hankel transform coincides with the 2D Fourier transform.

\section{Field propagation}\label{sec:propagation}

In this section, we show how to calculate the TST, ST, and TS contributions to the propagation functions $F$, defined in Eq.~(\ref{eq:F_definition}),
while the detailed calculations can be found in Appendix~\ref{app:F_terms}.
As noted, these functions carry all the geometrical properties of the sample+tip setup and the incident field, but do not include the phonon polarizability response of the medium, which we treated in the previous section.
The respective propagators for the TST, ST, and TS terms are \cite{cancado2014}
\begin{subequations}\label{eq:F_contributions}
    \begin{eqnarray}\label{eq:FTST}
        F_{ijk}^{\text{TST}} (\bsrho; \mbr_0, \omega_s,\omega)
        &=&
        \left[
            \sum_{n,n'}
            G^{\text{FF}}_{kn'}(\mbr_0,z\mbh{z};\omega_s)
            \frac{\omega_s^2}{\varepsilon_0 c^2} \alpha^{\text{tip}}_{nn'}
            G^{\text{NF}}_{nj}(z\mbh{z},\bsrho;\omega_s)
        \right]
        \notag\\
        &&\times \left[
            \sum_{l,l'}
            G^{\text{NF}}_{il'}(\bsrho,z\mbh{z};\omega)
            \frac{\omega^2}{\varepsilon_0 c^2} \alpha^{\text{tip}}_{ll'}
            E_l (z\mbh{z}, \omega)
        \right],
    \end{eqnarray}
    \begin{equation}\label{eq:FST}
        F_{ijk}^{\text{ST}} (\bsrho; \mbr_0, \omega_s,\omega)
        =
        G^{\text{FF}}_{kj}(\mbr_0,\bsrho;\omega_s)
        \left[
            \sum_{l,l'}
            G^{\text{NF}}_{il'}(\bsrho,z\mbh{z};\omega)
            \frac{\omega^2}{\varepsilon_0 c^2} \alpha^{\text{tip}}_{ll'}
            E_l (z\mbh{z}, \omega)
        \right],
    \end{equation}
    \begin{equation}\label{eq:FTS}
        F_{ijk}^{\text{TS}} (\bsrho; \mbr_0, \omega_s,\omega)
        =
        \left[
            \sum_{n,n'}
            G^{\text{FF}}_{kn'}(\mbr_0,z\mbh{z};\omega_s)
            \frac{\omega_s^2}{\varepsilon_0 c^2} \alpha^{\text{tip}}_{nn'}
            G^{\text{NF}}_{nj}(z\mbh{z},\bsrho;\omega_s)
        \right]
        E_i (\bsrho, \omega),
    \end{equation}
\end{subequations}
where $G^{\text{NF}}_{il'}(\bsrho,z\mbh{z};\omega)$ is a near field propagator from $z\mbh{z}$ (on tip axis) to $\bsrho$ (on sample plane), from polarization $i$ to $l'$,
at frequency $\omega$,
and analogously for the other $G$ functions
(e.g. $G^{\text{NF}}_{nj}(z\mbh{z},\bsrho;\omega_s)$ is the propagator from sample to tip, $\bsrho$ to $z\mbh{z}$),
with FF for a far field propagator.
Also note that we assumed that the tip response $\alpha^{\text{tip}}_{nn'}$ is the same for $\omega$ and $\omega_s$.
With this notation, regular Raman without the tip is
\begin{equation}\label{eq:FS}
    F_{ijk}^{\text{S}} (\bsrho; \mbr_0, \omega_s,\omega)
    =
    G^{\text{FF}}_{kj}(\mbr_0,\bsrho;\omega_s)
    E_i (\bsrho, \omega).
\end{equation}

We will assume that the tip responds much more strongly in the $z$ direction than in the sample plane directions and its response is the same for $\omega$ and $\omega_s$, that is,
$\alpha^{\text{tip}}_{nn'} (\omega) \approx 2\pi\varepsilon_0 r_{\text{tip}}^3 f_e \delta_{nz}\delta_{n'z}$,
where $r_{\text{tip}}$ is the tip radius and $f_e$ is a real enhancement factor.
This approximation was used in Ref.~\cite{cancado2014,publio2022}
and it is equivalent to the assumption that the tip emits radiation as a point dipole,
which has been experimentally verified as a valid approximation in Ref.~\cite{miranda2023}.
Then, up to a phase factor, the near field propagator in a medium with refractive index $n$ is \cite{novotny}
\begin{equation}\label{eq:GNF_text}
    G_{ij}^{\text{NF}} (\bsrho, z\mbh{z}; \omega)
    = \frac{c^2}{4\pi n^2\omega^2}
    \left[
        - \delta_{ij} \frac{1}{( \rho^2 + z^2 )^{3/2}}
        +\frac{3 r_i r_j}{( \rho^2 + z^2 )^{5/2}}
    \right],
\end{equation}
where $\delta_{ij}$ is the Kronecker delta,
$\mbr \equiv \bsrho + z\mbh{z}$,
with $\bsrho \equiv x\mbh{x} + y\mbh{y}$,
such that $r_x \equiv x$, $r_y \equiv y$ and $r_z \equiv z$,
and $\rho \equiv |\bsrho|$.
Note that interchanging either $i$ and $j$ or $\bsrho$ and $z\mbh{z}$ does not change the expression.
In comparison to Ref. \cite{cancado2014,publio2022}, we have to include the $z$ response of the sample, because we are open to $\nu > 0$ in Eq.~(\ref{eq:alpha_matrices}).

In our treatment, we will use the assumption that the radial component of the incident electric field in the sample+tip system can be neglected,
because the use of a focused radially polarized Gaussian beam yields a $z$ component at least one order of magnitude stronger than the radial one
in the vicinity of the tip
(see Fig.~\ref{fig:field_rz_plane} in Appendix~\ref{app:beam}).
This causes the TS contribution originating from the in-plane component of the incident field to be negligible in comparison with TST and ST, so it is safe to assume it is null.
It is possible that the TS contribution originating from the out-of-plane component is much larger than the in-plane one, and we include it in our treatment. In the case that it is small, thus being comparable to TS in-plane, the error in including one and not the other will not be significant, because both will be negligible in comparison with TST and ST.
It is worth noting that in Ref.~\cite{publio2022} the authors assume that the radially polarized field at the vicinity of the tip has an intensity of about 1/3 of that of the $z$ component, which would imply that the radially polarized TS component cannot be neglected.
In Appendix~\ref{app:beam} we assume a strong focus specified by an objective numerical aperture (NA) of 1.4, leading to the negligible radial component in the vicinity of the tip, as seen in Fig.~\ref{fig:field_rz_plane}.
For weaker focusing, for instance NA $\lesssim 1$,
the radial TS contribution might still be important.

In addition, since the $z$ component of the incident field does not vary significantly in the region around the tip, we will neglect its dependence on $\mbr$ and $z$.
We also take advantage of the cylindrical symmetry of the problem and write the far field propagator as
the sum of an axial $g_z^{\T{FF}}$ and a radial $g_\parallel^{\T{FF}}$ part,
\begin{equation}\label{eq:GFF}
    G^{\text{FF}}_{kj} = \delta_{kj} [ g^{\text{FF}}_z \delta_{jz} + g^{\text{FF}}_{\parallel} (1-\delta_{jz})].
\end{equation}
These factors indicate how much light with each polarization is collected from the scattering and directed towards the detectors.
If $g_z^{\T{FF}}$ and $g_\parallel^{\T{FF}}$ differ, it means that the system collects one polarization better than the other,
and this will depend on the particular geometry of the experimental setup.

The explicit forms of the $F$ functions of Eq.~(\ref{eq:F_contributions}) are calculated in Appendix~\ref{app:F_terms}, and
the full propagator is the sum of all the contributions,
\begin{equation}\label{eq:F_sum_of_contributions}
    F_{ijk} (\mb{q}; \mbr_0, \omega_s,\omega)
    =
    F_{ijk}^{\text{TST}} (\mb{q}; \mbr_0, \omega_s,\omega)
    + F_{ijk}^{\text{ST}} (\mb{q}; \mbr_0, \omega_s,\omega)
    + F_{ijk}^{\text{TS}} (\mb{q}; \mbr_0, \omega_s,\omega).
\end{equation}

The index associated with the detected field polarization is $k$, so if we group the components by this index, we can see which contributions interfere with each other.
Namely, TST and TS contain only a $k=z$ contribution (because they scatter from the tip last), while ST contributes with all components, $k\in\{x,y,z\}$ (because it scatters from the sample last).
This means that interference between contributions TST, ST and TS will appear in the scattered $z$ component, because they are scattered towards the detector in the same polarization,
\begin{subequations}
    \begin{equation}
        F_{ijz} =
        F_{ijz}^{\text{TST}} + F_{ijz}^{\text{ST}} + F_{ijz}^{\text{TS}},
    \end{equation}
    \begin{equation}
        F_{ijx} =
        F_{ijx}^{\text{ST}},
    \end{equation}
    \begin{equation}
        F_{ijy} =
        F_{ijy}^{\text{ST}}.
    \end{equation}
\end{subequations}
Note that in Ref. \cite{cancado2014}, in which there is no TS contribution and no sample $z$ response, the total contribution is just a sum of TST and ST intensities, without any interference, because there $F_{ijz} = F_{ijz}^{\text{TST}}$.

\section{TERS signal of particular phonon modes}\label{sec:ters_of_modes}

In the previous section, the propagation of the electromagnetic field in the TERS signal was studied independently of the sample properties.
The phonon symmetries, however, given for instance in the Raman tensors $\alpha^{(\eta)}$ of Eq.~(\ref{eq:alpha_matrices}),
select the polarization of the scattered light, so not all components of the tensor $F_{ijk}$ participate in the TERS signal of a given mode.
To understand which $i,j$ components of the tensor $F_{ijk}$ are probed by each mode, and in which combination,
we check its multiplication with the $\alpha^{(\eta)}$ for each mode,
\begin{subequations}\label{eq:F_sums}
    \begin{equation}
        \sum_{i,j} F_{ijk} \alpha_{ij}^{(A_1)}
        = \alphabar^{(A_1)}\left( F_{xxk} + F_{yyk} + \nu F_{zzk} \right),
    \end{equation}
    \begin{equation}
        \sum_{i,j} F_{ijk} \alpha_{ij}^{(E_{2g},x)}
        = \alphabar^{(E_{2g})}\left( F_{xyk} + F_{yxk} \right),
    \end{equation}
    \begin{equation}
        \sum_{i,j} F_{ijk} \alpha_{ij}^{(E_{2g},y)}
        = \alphabar^{(E_{2g})}\left( F_{xxk} - F_{yyk} \right).
    \end{equation}
\end{subequations}

All the tensorial properties of the TERS scattering of mode $\eta$ are contained in $\mc{J}^{(\eta)}$, appearing in Eq.~(\ref{eq:detected_signal_mcJ}).
Deriving the following final form for each mode from Eq.~(\ref{eq:F_sums}) is
done in Appendix~\ref{app:I_calculation},
using the results of Appendixes~\ref{app:F_terms}--\ref{app:bessel},
and we write the result as a function of $q$ and $z$ below,
\begin{subequations}\label{eq:Jeta_results}
    \begin{eqnarray}\label{eq:Jeta_A1}
        \tilde{\mc{J}}^{(A_1)}(q;z)
        =
        (2\pi)^3 |\alphabar^{(A_1)}|^2 |E_z (\omega)|^2
        &\Bigg\{&
        \left|g^{\text{FF}}_{z}\right|^2
        \Bigg|
        \left(
            \frac{r_{\text{tip}}^3 f_e}{2n^2}
        \right)^2
        \frac{9}{384}
        \left[ \frac{2 q^3 K_3(qz)}{z} - q^4 K_2(qz) \right]
        \notag\\
        &&+ \nu \left(
            \frac{r_{\text{tip}}^3 f_e}{2n^2}
        \right)^2
        \left[
            \frac{9}{384} q^4 K_4(qz)
            - \frac{6}{48} \frac{q^3 K_3(qz)}{z}
            + \frac{1}{8} \frac{q^2 K_2(qz)}{z^2}
        \right]
        \notag\\
        &&+
        2 \nu
            \left(
                \frac{r_{\text{tip}}^3 f_e}{2n^2}
            \right)
            q e^{-qz}
        \Bigg|^2
        \notag\\
        &&+
        \left|g^{\text{FF}}_{\parallel}\right|^2
        \left|
            \left(
                \frac{r_{\text{tip}}^3 f_e}{2n^2}
            \right)
            q e^{-qz}
        \right|^2
        \Bigg\}
        ,
    \end{eqnarray}
    \begin{eqnarray}\label{eq:Jeta_E2g}
        \tilde{\mc{J}}^{(E_{2g})}(q;z)
        =
        (2\pi)^3 |\alphabar^{(E_{2g})}|^2 |E_z (\omega)|^2
        &\Bigg\{&
        \left|g^{\text{FF}}_{z}\right|^2
        \left|
            \left(
                \frac{r_{\text{tip}}^3 f_e}{2n^2}
            \right)^2
            \left(\frac{9}{384}\right) q^4 K_2(qz)
        \right|^2
        \notag\\
        &&+
        2 \left|g^{\text{FF}}_{\parallel}\right|^2
        \left|
            \left(
                \frac{r_{\text{tip}}^3 f_e}{2n^2}
            \right)
            q e^{-qz}
        \right|^2
        \Bigg\},
    \end{eqnarray}
\end{subequations}
where we used $\alpha^{\text{tip}}_{zz} = 2\pi\varepsilon_0 r_{\text{tip}}^3 f_e$ \cite{cancado2014}
and $\mc{J}^{(E_{2g})}(q;z) = \mc{J}^{(E_{2g},x)}(q;z) + \mc{J}^{(E_{2g},y)}(q;z)$, assuming that $L_c$ is the same for the two branches of mode E$_{2g}$.
The first term in which $\nu$ appears in Eq.~(\ref{eq:Jeta_A1}) comes from the TST contribution in the A$_1$ mode,
and the second one is a sum of ST and TS contributions in this mode, which are equal in magnitude, leading to the factor 2 in the term.

Note that in Eq.~(\ref{eq:Jeta_results}) there are factors that are powers of $r_{\T{tip}}^3 f_e/(2n^2)$ defining the strength of each contribution to the TERS near field intensity.
The tip radius $r_{\T{tip}}$ will enter the tip-sample distance $z$, so it affects the result in more than one way,
but $f_e$ and $n$ only appear in the mentioned factor.
This means that they can be joined into a single parameter
\begin{equation}\label{eq:feeff}
    f_e^{\T{eff}} \equiv f_e/n^2,
\end{equation}
which behaves as an effective tip enhancement factor modified by the medium refractive index,
and the factors in Eq.~(\ref{eq:Jeta_results}) can be rewritten according to
$r_{\T{tip}}^3 f_e/(2n^2) \rightarrow r_{\T{tip}}^3 f_e^{\T{eff}}/2$.

\begin{figure*}[ht]
\begin{center}
    \includegraphics[width=1.\textwidth]{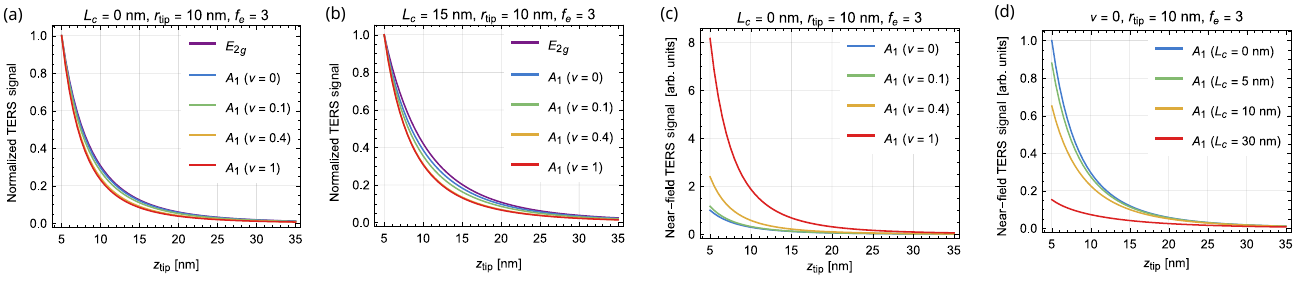}
\end{center}
    \caption{\label{fig:ters_nu}
        [(a) and (b)] Normalized TERS signal according to Eq.~(\ref{eq:detected_signal_mcJ}) and Eq.~(\ref{eq:Jeta_results}) for modes A$_1$ and E$_{2g}$ for a variety of $\nu$ values, as a function of the tip-sample distance,
        with fixed $L_c$, $r_{\text{tip}}$ and $f_e$.
        [(c) and (d)] Absolute value of the near field TERS signal according to Eqs.~(\ref{eq:detected_signal_mcJ}) and (\ref{eq:Jeta_results}) for mode A$_1$, where in (c) we show a variety of $\nu$ values, while in (d) curves for different $L_c$ are shown,
        for fixed $r_{\text{tip}}$ and $f_e$.
        The fixed values within each plot are written on top of each frame.
        According to Fig.~\ref{fig:scattering}, we call $z_{\text{tip}}$ the distance from the sample surface to the tip point, which is the quantity on the horizontal axis of the plots, such that $z = z_{\text{tip}} + r_{\text{tip}}$ is the value used in Eq.~(\ref{eq:Jeta_results}).}
\end{figure*}

From Eq.~(\ref{eq:detected_signal_mcJ}), the TERS signal of mode $\eta$ is the inner product between $\mc{J}^{(\eta)}$ and the $L_c$ dependent correlation function $\mc{L}^{(\eta)}$.
In Figs.~\ref{fig:ters_nu}~(a)--(c), we plot the TERS signal for modes A$_1$ and E$_{2g}$ for a variety of $\nu$ values (note that only A$_1$ depends on $\nu$), as a function of the tip-sample distance, as is done in tip approach experiments \cite{cancado2009,publio2022,nadas2025},
with fixed $L_c$, $r_{\text{tip}}$ and $f_e$, and write on the top of the plot the values used.
In Fig.~\ref{fig:ters_nu}~(d), we fix $\nu$, $r_{\text{tip}}$ and $f_e$, and plot the curves for a variety of $L_c$ values.
As in Fig.~\ref{fig:scattering}, we call $z_{\text{tip}}$ the distance from the sample surface to the tip point, which is the quantity in the horizontal axis of the plots, such that $z = z_{\text{tip}} + r_{\text{tip}}$ is the position of the dipole inside the tip that we use to model the field emitted by it.
In Figs.~\ref{fig:ters_nu}~(a) and (b), all the curves are normalized within themselves, with the reference at the closest point at which the tip gets to the sample, which we take at $z_{\text{tip}} = z_0 = 5 \text{ nm}$ [see Fig.~\ref{fig:scattering}].
In Figs.~\ref{fig:ters_nu}~(c) and (d), we compare the absolute intensity of the curves, and for that we take the A$_1$ near field TERS intensity with $\nu=0$ and $L_c=0\T{ nm}$ at $z_{\text{tip}} = 5 \text{ nm}$ as unity, and we do not plot E$_{2g}$ because its intensity in relation to A$_1$ can change from material to material.

We see in Figs.~\ref{fig:ters_nu}~(a) and (b) that increasing $\nu$ increases the distance between the A$_1$ and E$_{2g}$ curves, for both $L_c = 0 \text{ nm}$ and $L_c = 15 \text{ nm}$.
It does not act alone, though, since a higher $L_c$ yields a higher impact of $\nu$.
Most strikingly, in Fig.~\ref{fig:ters_nu}~(c), an increase in $\nu$ dramatically increases the intensity of the near field TERS signal.
This means that the $z$-axis sample Raman response can have a significant impact on the TERS enhancement.
Note in Fig.~\ref{fig:ters_nu}~(d) that as $L_c$ increases the absolute value of the near field TERS intensity decreases
(the plot displays only the A$_1$ mode, but for E$_{2g}$ the same behavior appears).
For the E$_{2g}$ mode this is expected due to the scattering destructive intereference arising from the phonon symmetry,
but it seems counter-intuitive for the A$_1$ mode, which displays constructive interference instead~\cite{cancado2014}, hinted in the Raman tensors of Eq.~(\ref{eq:alpha_matrices})
and Eq.~(\ref{eq:F_sums}).
This behavior follows from the norm of $\mc{L}$ being proportional to $1/L_c$,
together with the fact that the TERS signal can be put in the form of an inner product with $\mc{L}$,
as mentioned in Sec.~\ref{sec:polarizability}.
Detailed discussions on the influence of the physical parameters in the TERS measurements follow in Sec.~\ref{sec:discussion}.

\section{Discussion}\label{sec:discussion}

The discussion is split into four parts, Sec.~\ref{sec:normalized} to discuss the normalized TERS measurements, as those in Figs.~\ref{fig:ters_nu}~(a) and (b),
Sec.~\ref{sec:enhancement} to discuss enhancement measurements, affected by the absolute intensity of the near field, as in Figs.~\ref{fig:ters_nu}~(c) and (d),
Sec.~\ref{sec:graphene} to address the implications of the out-of-plane response to the fitting of graphene measurements,
and Sec.~\ref{sec:correlation_functions} to examine other possible correlation functions for our model.

\subsection{Normalized TERS measurements}\label{sec:normalized}

In normalized TERS measurements, as those in Figs.~\ref{fig:ters_nu}~(a) and (b),
it is the distance between the E$_{2g}$ and A$_1$ curves that testifies the difference in steepness between the tip approach curves of the two modes \cite{beams2014,cancado2014,publio2022}.
With that in mind, we construct a figure of merit based on the integral of the normalized tip approach curve of each mode between points $z_1$ and $z_2$,
\begin{equation}\label{eq:W}
    \mc{W}^{(\eta)}(z_1,z_2)
    \equiv
    \frac{
        \displaystyle \int_{z_1}^{z_2}
        S^{(\eta)}(\mb{r}_0,\omega_s;z)
        dz
    }
    {S^{(\eta)}(\mb{r}_0,\omega_s;z_1)}
    .
\end{equation}
We always pick $z_1 = z_0+r_{\T{tip}}$, and a $z_2$ large enough so the two curves practically match each other,
in particular $z_2 = z_0+r_{\T{tip}}+ 100\T{ nm}$.
The figure of merit is the difference between the integrals of the two modes,
\begin{equation}\label{eq:deltaW}
    \Delta\mc{W}
    \equiv
    \mc{W}^{(E_{2g})}(z_1,z_2)
    -\mc{W}^{(A_{1})}(z_1,z_2),
\end{equation}
or equivalently the area between the two curves,
being positive when E$_{2g}$ is most of the times above A$_1$, which is the standard situation.
The larger $\Delta\mc{W}$, the more noticeable is the difference between the E$_{2g}$ and A$_1$ curves.

\begin{figure*}[t]
\begin{center}
    \includegraphics[width=1.0\textwidth]{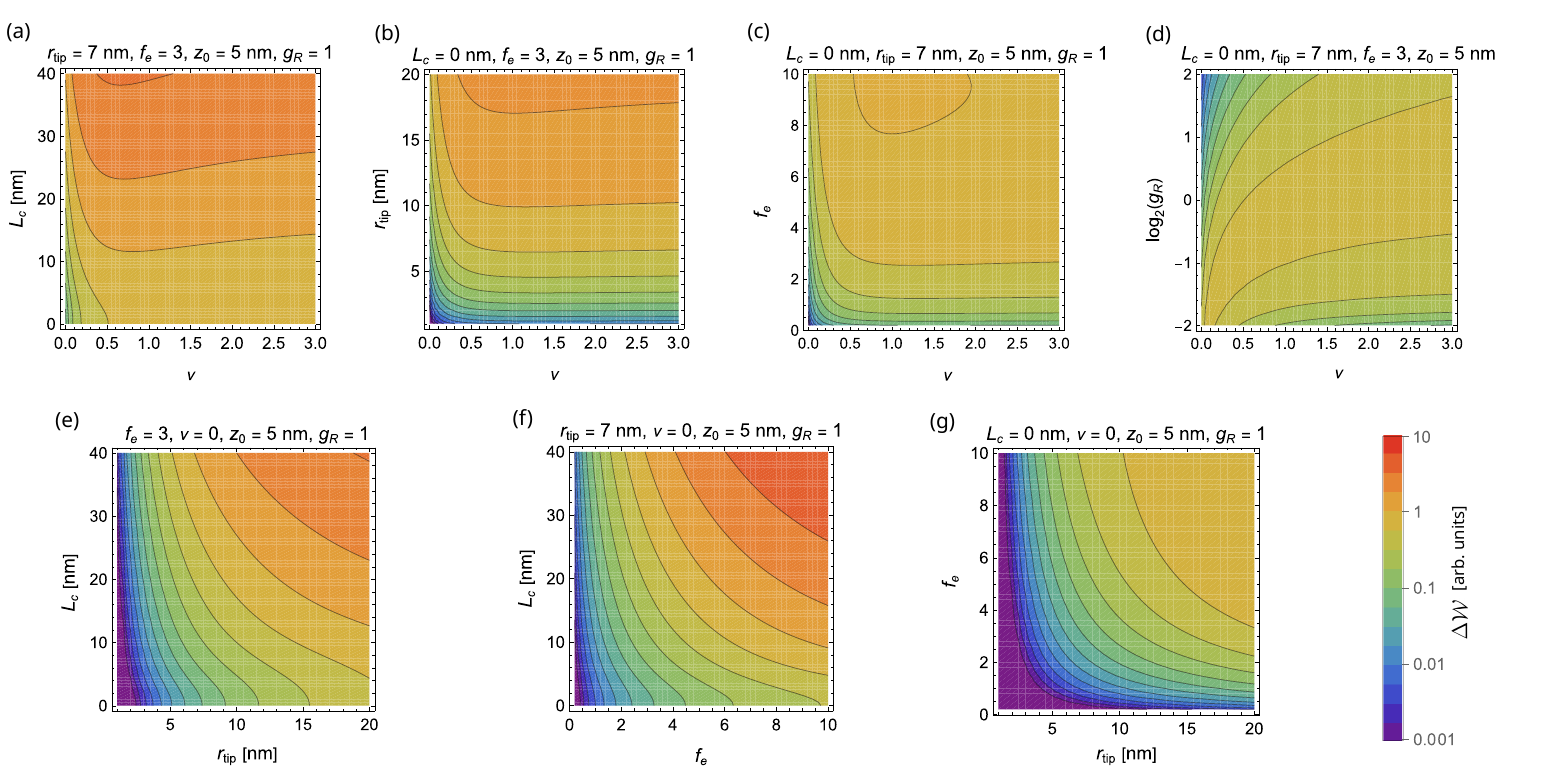}
\end{center}
    \caption{\label{fig:params} Plots of the figure of merit $\Delta\mc{W}$, defined by Eqs.~(\ref{eq:W}) and (\ref{eq:deltaW}),
    which quantifies the difference under the normalized tip approach curves between modes A$_1$ and E$_{2g}$.
    In each of the plots, a pair of physical parameters in the model is varied, while the others are fixed.
    The varied parameters encompass the coherence length $L_c$, the material out-of-plane relative response $\nu$, the tip radius $r_{\T{tip}}$, the tip response factor $f_e$,
    and the ratio of the far field propagation factors $g_{\T{R}} = g_z^{\T{FF}}/g_{\parallel}^{\T{FF}}$, of which we plot $\log_2(g_\T{R})$.
    We use $z_0 = 5 \T{ nm}$ in all plots,
    and the values for the fixed parameters in each plot appear written above each one respectively.
    We use the same $\log_{10}$ color scale for $\Delta\mc{W}$ across all plots, and the contours are separated by $10^{1/4}$ [arb. units] in all of them.
    }
\end{figure*}

In Fig.~\ref{fig:params}, we plot 2D maps of $\Delta\mc{W}$ for different pairs of parameters in our model,
Eqs.~(\ref{eq:detected_signal_mcJ}) and (\ref{eq:Jeta_results}),
namely the coherence length $L_c$, the material out-of-plane relative response $\nu$, the tip radius $r_{\T{tip}}$, the tip response factor $f_e$,
and the ratio of the far field propagation factors $g_{\T{R}} \equiv g_z^{\T{FF}}/g_{\parallel}^{\T{FF}}$.
The parameters $L_c$ and $\nu$ are characteristics of the material,
while $r_{\T{tip}}$, $f_e$ and $g_\T{R}$ vary with the experimental settings.
The values for the fixed parameters in all plots are $L_c = 0 \T{ nm}$, $\nu=0$, $r_{\T{tip}} = 7 \T{ nm}$, $f_e = 3$, and $g_{\T{R}} = 1$, except when they are varied, with $z_0 = 5 \T{ nm}$ in all plots without exception.
In Eq. (\ref{eq:Jeta_results}) there is also a dependence on the refractive index $n$ of the medium between the tip and sample,
but with the definition of Eq.~(\ref{eq:feeff}),
the refractive index enters as a modification of $f_e$, leading to an effective $f_e^{\T{eff}}$.
Therefore we do not plot the $n$ dependence explicitly,
and the effect of $n$ is obtained by interpreting $f_e$ as $f_e^{\T{eff}}$.
Notice that $g_z^{\T{FF}}$ and $g_{\parallel}^{\T{FF}}$, and consequently $g_\T{R}$, depend on how much of the scattered light in each polarization the optical system is able to collect and direct towards the detectors.
Also, because $g_\T{R}$ is a ratio, we plot $\log_2(g_\T{R})$ to have a balanced representation of $g_\T{R} > 1$ and $g_\T{R} < 1$.
Importantly, we use the same $\log_{10}$ color scale for $\Delta\mc{W}$ across all plots of Fig.~\ref{fig:params}, and the contours are separated by $10^{1/4}$ [arb. units] in all of them.

At first glance, it is noticeable that the plots with $L_c$ [Fig.~\ref{fig:params}~(a), \ref{fig:params}~(e) and \ref{fig:params}~(f)] are the ones with the highest values of $\Delta\mc{W}$,
meaning that a large $L_c$ is the single most determinant factor to a large separation between the two curves.
This is attributed to a larger destructive interference in the sample scattering pattern of the E$_{2g}$ mode when the coherence length is large \cite{cancado2014,publio2022}.

When $L_c = 0 \T{ nm}$, the highest values of $\Delta\mc{W}$ occur for $\nu > 0.5$ combined with a large $r_{\T{tip}}$ [Fig.~\ref{fig:params}~(b)].
Also note that, when varying $\nu$, $\Delta\mc{W}$ plateaus from $\nu \approx 1$ onwards,
and for $\nu > 1$ there is just a slight decrease in $\Delta\mc{W}$
for all variables except $g_R$ [Figs.~\ref{fig:params}~(a)--(d)].
For $\log_2(g_{\T{R}}) \approx 0$ ($g_{\T{R}} \approx 1$) with high $\nu$ it might seem that $\nu > 1$ still has a significant impact on the growth of $\Delta\mc{W}$ [Fig.~\ref{fig:params}~(d)],
but note that in this plot $\Delta\mc{W} < 1$ [arb. units] everywhere, being lower than all other $\nu$ plots [Figs.~\ref{fig:params}~(a)--(d)].
We interpret this as $g_{\T{R}}$ being the least important parameter in the variation of $\Delta\mc{W}$.

When it comes to the refractive index $n$, its variation is here interpreted as a rescaling of $f_e$ in Eq.~(\ref{eq:Jeta_results}) according to Eq.~(\ref{eq:feeff}).
For instance, in a medium where $n=3$ one has $f_e^{\T{eff}}= f_e/9$,
so if $f_e = 4$ in a TERS setting with air between sample and tip,
we would have $f_e^{\T{eff}}= 0.44$.
As a rule of thumb, the larger is $n$, the smaller is $f_e^{\T{eff}}$, and this means a smaller $\Delta\mc{W}$ [Figs.~\ref{fig:params}~(c), (f) and (g)].
The intuition behind this is that a large $n$ causes the optical path of the field in the region between the sample and the tip to be large, so the near field becomes weaker and the difference between the normalized curves decreases significantly.

\subsection{TERS enhancement measurements}\label{sec:enhancement}

The effects of $\nu$ and $g_{\T{R}}$ are much more pronounced in non-normalized measurements, because they can significantly increase the scattered field signal for small tip-sample distances, as seen in Eq.~(\ref{eq:Jeta_results}) and Fig.~\ref{fig:ters_nu}~(c).
When trying to model this kind of measurement, we need additional parameters, namely the scattered signal when the tip is distant from the sample (``tip up'' setting),
corresponding to the scattered Raman far field $S_{\T{FF}}^{(\eta)}$ for mode $\eta$.
We need to specify $S_{\T{FF}}^{(\eta)}$ in relation to the signal for a small tip-sample distance (``tip down'' setting) for each mode, that is including the near field,
and we also need to compare the values between E$_{2g}$ and A$_1$ modes.
To do that, we use
\begin{equation}\label{eq:ffnf}
    S_{\T{total}}^{(\eta)}(z) =
    S_{\T{NF}}^{(\eta)}(z)
    +
    S_{\T{FF}}^{(\eta)}
    ,
\end{equation}
where $S_{\T{NF}}(z)$ is the scattered signal calculated in Eq.~(\ref{eq:detected_signal_mcJ}), which depends on the tip-sample distance $z$,
and $S_{\T{FF}}(z)$ is the scattered far field signal, which is independent of $z$.

The insightful experimental physical quantity in this framework is the TERS spectral enhancement,
which is the intensity of the scattered Raman signal when the tip is engaged in the sample in relation to when it is far from the sample (intensity at ``tip down'' divided by intensity at ``tip up''), defined by
\begin{equation}\label{eq:enhancement}
    E^{(\eta)} \equiv
    \frac{S_{\T{total}}^{(\eta)}(z_0+r_{\T{tip}})}
    {S_{\T{total}}^{(\eta)}(z_0+r_{\T{tip}}+z_F)},
\end{equation}
with $z_F \gtrsim 10 z_0$.
When $z_F$ is large enough, $S_{\T{total}}^{(\eta)}(z_0+r_{\T{tip}}+z_F) \approx S_{\T{FF}}^{(\eta)}$, so we can write
\begin{equation}
    E^{(\eta)} \approx
    \frac{S_{\T{NF}}^{(\eta)}(z_0+r_{\T{tip}})}
    {S_{\T{FF}}^{(\eta)}}
    +1
    .
\end{equation}

\begin{figure}[ht]
\begin{center}
    \includegraphics[width=.47\textwidth]{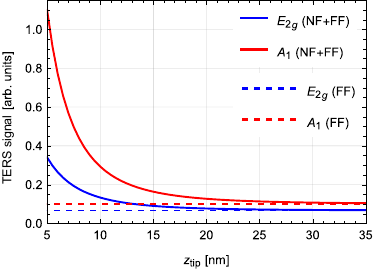}
\end{center}
    \caption{\label{fig:ffnf} Construction of the total TERS signal (solid lines), according to Eq.~(\ref{eq:ffnf}),
    comprised of the contribution of the near field TERS (NF), calculated with Eq.~(\ref{eq:detected_signal_mcJ}),
    and a far field (FF) signal constant along $z_{\T{tip}}$ (dashed lines).
    In the plot we use, for the A$_1$ mode (red lines), $S_{\T{FF}}^{(A_1)} = (1/10) S_{\T{NF}}^{(A_1)}(z_{0}+r_{\T{tip}})|_{L_c=0}$, with $z_0 = 5 \T{ nm}$ and $r_{\T{tip}} = 7 \T{ nm}$,
    and for E$_{2g}$ (blue lines) we use $S_{\T{FF}}^{(E_{2g})} = (2/3) S_{\T{FF}}^{(A_1)}$.
    For both modes we use $L_c = 0 \T{ nm}$, $f_e = 3$, $\nu = 1$ and $g_R = 1$.
    }
\end{figure}

In Fig.~\ref{fig:ffnf} we plot the tip approach signal $S_{\T{total}}^{(\eta)}(z)$ for the two modes in solid lines, and in dashed lines we plot the far field signal $S_{\T{FF}}^{(\eta)}(z)$ that we have used.
The intensities used are $S_{\T{FF}}^{(A_1)} = (1/10) S_{\T{NF}}^{(A_1)}(z_{0}+r_{\T{tip}})|_{L_c=0}$, with $z_0 = 5 \T{ nm}$ and $r_{\T{tip}} = 7 \T{ nm}$,
and for E$_{2g}$ we use $S_{\T{FF}}^{(E_{2g})} = (2/3) S_{\T{FF}}^{(A_1)}$,
which yield $E^{(E_{2g})} = 5.0$ and $E^{(A_1)} = 10.5$.
These values are chosen because they are close to the ones appearing in an MoSe$_2$ experiment~\cite{guimaraes2025}.
The other parameters of the plot in Fig.~\ref{fig:ffnf} are $L_c = 0 \T{ nm}$, $f_e = 3$, $\nu = 1$ and $g_R = 1$.

\begin{figure*}[ht]
\begin{center}
    \includegraphics[width=1.\textwidth]{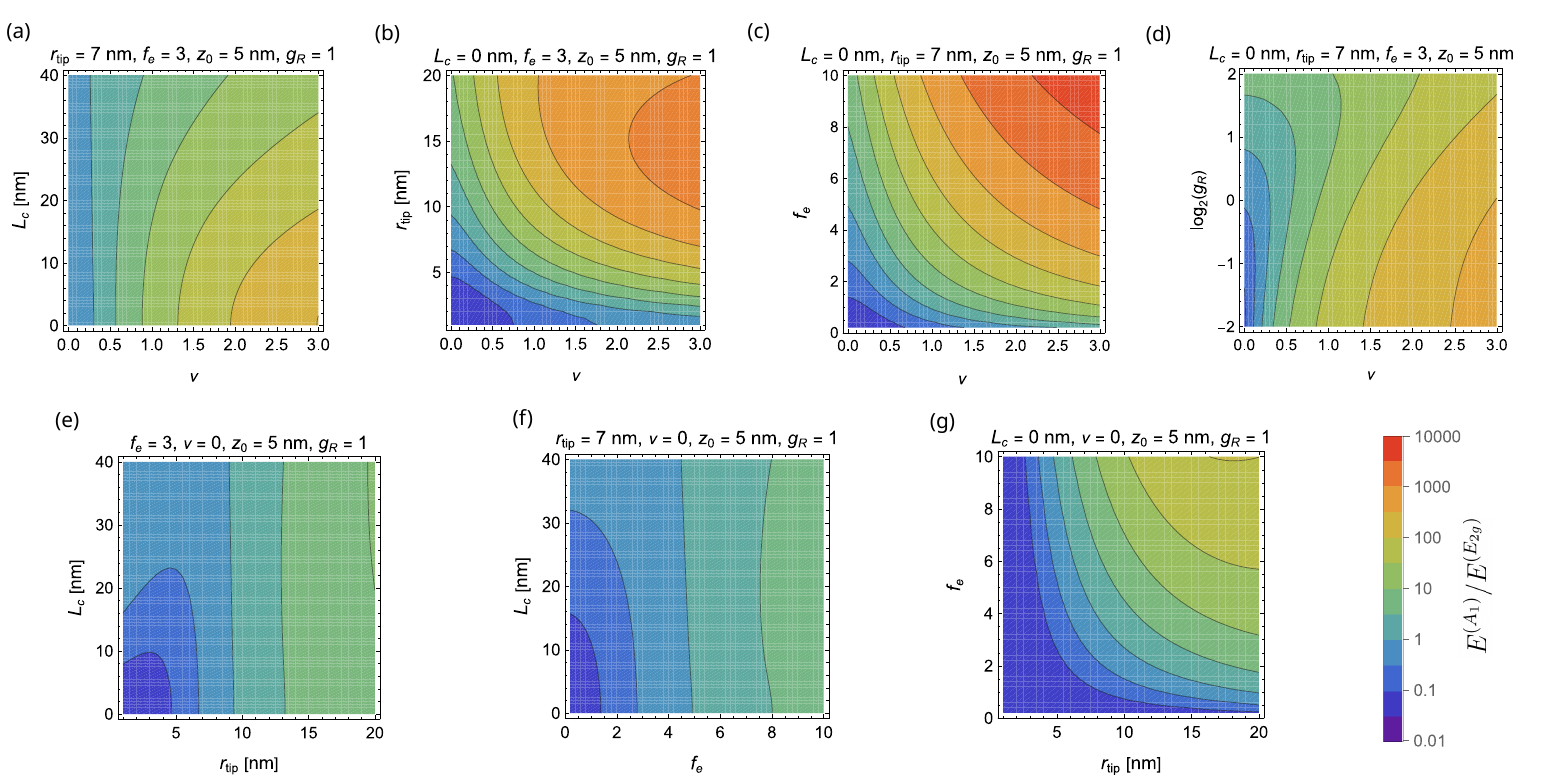}
\end{center}
    \caption{\label{fig:enhancement} Plots of the enhancement ratio $E^{(A_1)}/E^{(E_{2g})}$,
    constructed from the enhancements defined in Eq.~(\ref{eq:enhancement}).
    In each of the plots, a pair of physical parameters in the model is varied, while the others are fixed.
    The varied parameters encompass the coherence length $L_c$, the material out-of-plane relative response $\nu$, the tip radius $r_{\T{tip}}$, the tip response factor $f_e$,
    and the ratio of the far field propagation factors $g_{\T{R}} = g_z^{\T{FF}}/g_{\parallel}^{\T{FF}}$, of which we plot $\log_2(g_\T{R})$.
    In all plots we use $z_0 = 5 \T{ nm}$,
    the far field intensities for the A$_1$ mode $S_{\T{FF}}^{(A_1)} = (1/10) S_{\T{NF}}^{(A_1)}(z_{0}+r_{\T{tip},0})|_{L_c=0}$,
    where $r_{\T{tip},0} = 7 \T{ nm}$,
    and for E$_{2g}$ we use $S_{\T{FF}}^{(E_{2g})} = (2/3) S_{\T{NF}}^{(A_1)}$.
    The values for the fixed parameters in each plot appear written above each one respectively.
    We use the same $\log_{10}$ color scale for $E^{(A_1)}/E^{(E_{2g})}$ across all plots, and the contours are separated by $10^{1/2}$ in all of them.
    }
\end{figure*}

The figure of merit to relate the two modes is the ratio $E^{(A_1)}/E^{(E_{2g})}$ of the enhancement of the A$_1$ mode to that of E$_{2g}$, which for the plot in Fig.~\ref{fig:ffnf} is $E^{(A_1)}/E^{(E_{2g})} = 2.1$.
In Fig.~\ref{fig:enhancement} we plot $E^{(A_1)}/E^{(E_{2g})}$ with varying parameters, analogously to what we did to $\Delta\mc{W}$ in Fig.~\ref{fig:params}.
We use the same far field Raman signal of the A$_1$ mode in all plots, namely $S_{\T{FF}}^{(A_1)} = (1/10) S_{\T{NF}}^{(A_1)}(z_{0}+r_{\T{tip},0})|_{L_c=0}$, with $z_0 = 5 \T{ nm}$ and $r_{\T{tip},0} = 7 \T{ nm}$,
and for E$_{2g}$ we use $S_{\T{FF}}^{(E_{2g})} = (2/3) S_{\T{NF}}^{(A_1)}$,
as in Fig.~\ref{fig:ffnf}.
These serve just as reference values, not playing a significant role in the tendencies shown in Fig.~\ref{fig:enhancement}.

As we expected, in the plots of Fig.~\ref{fig:enhancement} it is evident that a large $\nu$ implies a large $E^{(A_1)}/E^{(E_{2g})}$,
as the maximum enhancement ratio in the $\nu$-dependent plots [Figs.~\ref{fig:enhancement}~(a)--(d)] is always larger than in the $\nu=0$ plots [Figs.~\ref{fig:enhancement}~(e)--(g)].
In addition, $f_e$ is the most important factor which, associated with a large $\nu$, leads to a large enhancement ratio [Fig.~\ref{fig:enhancement}~(c)], because it increases the TERS intensity, as evident in Eq.~(\ref{eq:Jeta_results}).
The tip radius $r_{\T{tip}}$ is also important, as it follows $f_e$ in Eq.~(\ref{eq:Jeta_results}), but since it also means a higher $z = z_0+r_{\T{tip}}$ distance, at some point its increase starts to lower $E^{(A_1)}/E^{(E_{2g})}$ [Fig.~\ref{fig:enhancement}~(b)].

Note that increasing $L_c$ leads the enhancement ratio $E^{(A_1)}/E^{(E_{2g})}$ towards unity, that is, a balanced enhancement between the two modes.
In Fig.~\ref{fig:enhancement}~(a), where for large $\nu$ the ratio is larger than unity, increasing $L_c$ decreases the enhancement ratio,
while in Fig.~\ref{fig:enhancement}~(e) and \ref{fig:enhancement}~(f), where the ratio is smaller than unity for small $r_{\T{tip}}$ and $f_e$, increasing $L_c$ increases the ratio.
This is explained by the behavior of our correlation function model, as presented in Sec.~\ref{sec:polarizability} and seen in Fig.~\ref{fig:ters_nu}~(d), that causes the TERS near field intensity to decrease as $L_c$ increases [in Fig.~\ref{fig:ters_nu}~(d) only mode A$_1$ is shown, but the E$_{2g}$ behavior is analogous].
Because the far field is fixed and independent of $L_c$, if the near field decreases for both A$_1$ and E$_{2g}$ modes, then both enhancements decrease,
causing the enhancement ratio $E^{(A_1)}/E^{(E_{2g})}$ to get closer to unity.

There are regions in the parameter space in which the ratio is smaller than unity,
meaning that the E$_{2g}$ enhancement is larger than A$_1$.
These regions are those for very small parameters, and appear more strikingly in Fig.~\ref{fig:enhancement}~(g) in which at least three of the parameters $f_e$, $r_{\T{tip}}$, $L_c$ and $\nu$ are small at the same time,
suppressing the signal of A$_1$ more than of E$_{2g}$.

A word must also be said concerning the tip radius as an enhancement factor.
In the model (see Sec.~\ref{sec:propagation} and Ref.~\cite{cancado2014}), the tip is a spherical body which originates a dipole emission pattern in response to the external electric field~\cite{jackson}.
This means that the response of the tip depends on its volume, which is where the $r_{\T{tip}}^3$ factors in Eq.~\ref{eq:Jeta_results} come from.
Therefore, even though a larger tip means that the distance of the oscillating effective dipole inside the tip is farther from the sample,
this can be compensated by a stronger response due to the larger dipole volume,
and thus a large $r_{\T{tip}}$, in the current model, can be associated with a larger enhancement.
Note however that there is a limit to this,
and for some parameters increasing $r_{\T{tip}}$ starts to decrease the enhancement,
as seen in Fig.~\ref{fig:enhancement}~(b) for large $\nu$.

It is important that $\nu$ should be apparent only in TMDs, which have a strong out-of-plane Raman response $\nu \sim 1$, in contrast to the weak graphene response $\nu \ll 1$.

\subsection{Graphene TERS experiments}\label{sec:graphene}

\begin{figure}[t]
\begin{center}
    \includegraphics[width=.36\textwidth]{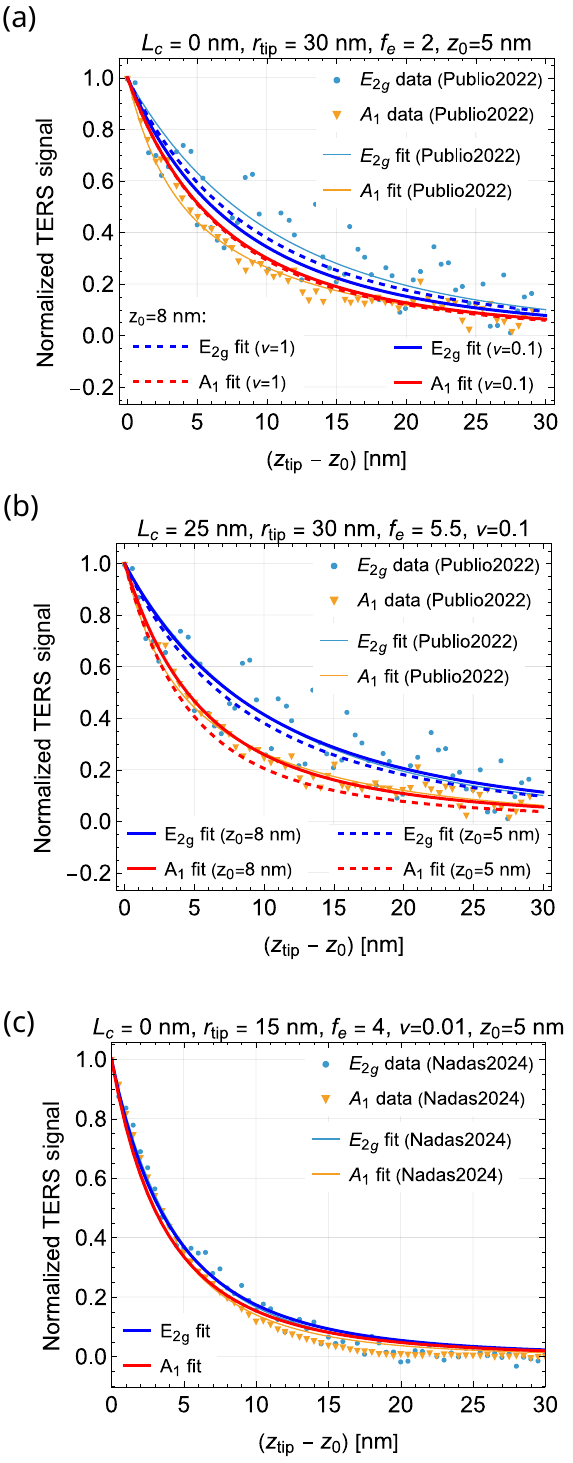}
\end{center}
    \caption{\label{fig:publionadas} Comparison between experimental TERS tip approach data collected in Ref.~\cite{publio2022} and \cite{nadas2025},
        the theoretical fitting of the original articles, and the fitting according to our theory, Eqs.~(\ref{eq:detected_signal_mcJ}) and (\ref{eq:Jeta_results}).
        Across all plots, light blue dots represent E$_{2g}$ mode data, inverted orange triangles are A$_1$,
        a light blue (orange) solid line is
        a fit with our theory using the fitting parameters of the original articles for E$_{2g}$ (A$_1$) mode,
        and a dark blue (red) solid line is the fit for E$_{2g}$ (A$_1$) mode with new parameters.
        In (a), data from Ref.~\cite{publio2022} are fit
        with $L_c=0$, both modes with $\nu=0.1$ and $z_0=5$ nm (solid line),
        and with $\nu=1$ and $z_0=8$ nm (dashed line).
        In (b) we use $L_c = 25$ nm, with $z_0=8$~nm (solid line) and $z_0=5$~nm (dashed line).
        In (c) data from Ref.~\cite{nadas2025} are fit with $L_c=0$ and a negligible $\nu=0.01$.
        The remaining parameters used in the fittings are written on top of each respective plot.
    }
\end{figure}

The discussion in Sec.~\ref{sec:normalized} leads to the prediction that a very large separation between E$_{2g}$ and A$_1$ normalized tip approach curves will inevitably need $L_c > 0$ to be modelled.
Curves with a smaller separation, on the other hand, can be modelled with $L_c = 0$ as long as $\nu$, $r_{\T{tip}}$, $f_e$, or some combination of them are large.
In Fig.~\ref{fig:publionadas} we show an example of this on data from Ref.~\cite{publio2022} [Fig.~\ref{fig:publionadas}~(a) and \ref{fig:publionadas}~(b)] and Ref.~\cite{nadas2025} [Fig.~\ref{fig:publionadas}~(c)], which perform tip approach TERS experiments with graphene.
In the three plots, light blue dots represent the normalized tip approach experimental data from E$_{2g}$ mode, orange triangles are data from A$_{1}$ mode,
and the light blue and orange solid thin lines are a fit using our own model,
but with the parameters provided in the respective original papers, all with $L_c > 0$.
The original parameters for Ref.~\cite{publio2022}
are $L_c = 40 \T{ nm}$, $r_{\T{tip}}=15\T{ nm}$ and $f_e=4$,
and those for Ref.~\cite{nadas2025}
are $L_c = 13 \T{ nm}$ for mode E$_{2g}$, $L_c = 10 \T{ nm}$ for mode A$_1$, $r_{\T{tip}}=8\T{ nm}$ and $f_e=2$,
and with $\nu=0$, $g_R=1$ and $z_0=5\T{ nm}$ for both.
It should be noted that in the data from Ref.~\cite{publio2022} the two mode curves are much more separated than those from Ref.~\cite{nadas2025},
so the first needs a much larger $L_c$ in their fit than the latter,
in accordance with our conclusion that a large $L_c$ is needed for widely separated curves.

To investigate the necessity of a nonzero TERS coherence length in the explanation of the experimental results, we first try to fit the data from Ref.~\cite{publio2022} with $L_c=0$.
These are the thicker solid ($\nu = 0.1$) and dashed ($\nu = 1$) blue (E$_{2g}$) and red (A$_1$) curves in Fig.~\ref{fig:publionadas}~(a).
As shown in the plot, $\nu = 0.1$ cannot reach the measured separation between the curves when $L_c = 0$.
If one goes to $\nu = 1$,
the curve separation can be enlarged significantly,
while the smallest tip-sample separation $z_0$ has to be set to 8~nm for the appropriate curve decay,
but this is still not enough.
On top of that, $\nu=1$ is unphysical, since for graphene $\nu \ll 1$.

In Fig.~\ref{fig:publionadas}~(b), we abandon the restriction on $L_c = 0$.
Our results of Fig.~\ref{fig:params} can be used as a guide to a fit similar to that of Ref.~\cite{publio2022}, but with a smaller $L_c$.
The plots in Fig.~\ref{fig:params}~(e) and \ref{fig:params}~(f) indicate that the same $\Delta\mc{W}$ can be achieved, given a decrease in $L_c$,
by an increase in $r_{\T{tip}}$ or $f_e$, and we increase both, as well as set $\nu=0.1$.
However, our figure of merit only captures the area between the curves, and not the absolute position of the curves, associated with the decay rate in the tip approach.
We are able to keep a similar $\Delta\mc{W}$ as in Ref.~\cite{publio2022} for $L_c = 25$~nm if we use a larger $r_{\T{tip}} = 30$~nm and $f_e = 5.5$,
but to place the curves in the appropriate position, we must use a larger $z_0 = 8$~nm (solid lines).
With $z_0 = 5$~nm (dashed lines) the curves decay too fast and pass below the data, and increasing $z_0$ raises the curves, since it is probing a region with a weaker near field.
This means that the results of Ref.~\cite{publio2022} can also be explained by a model with a significantly smaller coherence length than assumed in their article,
as long as there are strong enhancing factors present, such as a larger $r_{\T{tip}}$, $f_e$, and $\nu$.

Importantly, note that the engaging distance $z_0$ may be seen as a restriction on the parameter range,
in the sense that $L_c = 20$~nm can only fit the data if $z_0 > 10$~nm, which is incompatible with the TERS setting used in the experiment.
It must be stressed that a large $L_c$ making the curves less steep is in accordance with the increase in $L_c$ causing a decrease in the overall TERS signal of both modes, discussed in Sec.~\ref{sec:ters_of_modes}.
So $L_c$ and $z_0$ go together in decreasing the steepness of the curves, such that to fit the same data with a smaller $L_c$, an increase in $z_0$ is necessary in compensation.
In a nutshell, the wide separation between the curves demands some combination of large enhancing factors, be it $r_{\T{tip}}$, $f_e$ or $\nu$,
while the restriction $z_0 \leq 8$~nm sets $L_c \gtrsim 25$~nm.
The more one knows the experimental parameters, the tighter the restriction on a measurement of $L_c$.

In Fig.~\ref{fig:publionadas}~(c), with data from Ref.~\cite{nadas2025},
because the curves are close, we can model them with $L_c = 0$, namely with $\nu = 0.01$ (which is indistinguishable from the same curve with $\nu = 0$).

We can conclude from this that the TERS measurements with the graphene sample from Ref.~\cite{publio2022} need $L_c > 0$ even in our new model that includes the out-of-plane response $\nu$,
while measurements in the sample from Ref.~\cite{nadas2025} do not and can be explained with $L_c = 0$.
This behavior might have several causes which we cannot infer from the tip approach curves alone, possibly the influence of local doping~\cite{nadas2023}, or some other feature affecting the coherence length.
The data from Ref.~\cite{nadas2025}, shown in Fig.~\ref{fig:publionadas}~(c), are the average of many experiments, and it thus shows a smaller fluctuation,
but we believe that the difference is too large to stem from this fact alone.
In either case, it seems that $\nu$ can be neglected in fittings of TERS measurements on graphene.

\subsection{Spatial correlation functions}\label{sec:correlation_functions}

At the end of Sec.~\ref{sec:ters_of_modes} we mentioned that, rather counter-intuitively, the near field TERS signal decreases with the increase of the correlation length $L_c$,
which follows from the norm of $\mc{L}$ being proportional to $1/L_c$,
while the TERS signal can be put in the form of an inner product with $\mc{L}$.
One could imagine that this is particular of our choice for a Gaussian $\mc{L}$,
and try to find other candidates for the correlation function.
We may expect that, on physical grounds, a good candidate must approach the Dirac delta function in the limit of $L_c\rightarrow 0$,
that is, its integral must evaluate to one.
For instance, the candidate $\mathcal{L}(\rho;L_c) = 2(\pi L_c)^{-2} \T{sinc}[(\rho/L_c)^2]$
has norm $\sqrt{2}/(\pi L_c)$, showing the same behavior as the Gaussian.
A candidate like a Lorentzian function, $\mathcal{L}(\rho;L_c) = (\pi L_c)^{-1} [(\rho/L_c)^2+1]^{-1}$,
does have a constant norm, independent of $L_c$, meaning that under this kind of correlation changing $L_c$ would not have any effect on the TERS signal.
It however fails in fulfilling the unit integral condition,
thus not working as a Dirac delta in the $L_c\rightarrow 0$ limit.
The integral of the Lorentzian function in fact diverges, as a consequence of its $\rho^{-2}$ decay for large $\rho$,
in opposition to the exponential decay of the Gaussian that limits its integral
(the sinc case is more complicated, because it oscillates between positive and negative values).
Thus, for a fixed $L_c$, a Lorentzian correlation function eventually exceeds the Gaussian one even if it starts smaller at $\rho=0$,
which we interpret as the first having longer correlations than the second.

To make a more general statement, let us consider a general cylindrical correlation function
\begin{equation}
    \mathcal{L}(\rho;L_c) =
    N(L_c)
    f(\rho/L_c),
\end{equation}
where $f(\rho/L_c)$ is the radial function and $N(L_c)$ is a normalization that will ensure that its integral is unit.
The integral of $\mathcal{L}$ is
\begin{equation}
    \int_{\mathbb{D}}
        \mathcal{L}(\rho;L_c)
        d^2\bsrho
    =
    2\pi
    N(L_c)
    \int_0^\infty
        f(\rho/L_c)
        \rho d\rho
    ,
\end{equation}
and by defining the integration variable $u\equiv \rho/L_c$, to get an integral independent of $L_c$, we find
\begin{equation}\label{eq:integral_L_u}
    \int_{\mathbb{D}}
        \mathcal{L}(\rho;L_c)
        d^2\bsrho
    =
    2\pi
    N(L_c)
    L_c^2
    \int_0^\infty
        f(u)
        u du.
\end{equation}
The condition of unit integral is translated to a condition that the integral is independent of $L_c$,
and, from Eq.~(\ref{eq:integral_L_u}), $N(L_c) \propto L_c^{-2}$ is a necessary condition for this, as in the Gaussian, Eq.~(\ref{eq:Lc}), and sinc definitions.
Note that the Lorentzian candidate has $N(L_c) \propto L_c^{-1}$ and violates the condition.

We now calculate the norm, which with the same $u\equiv \rho/L_c$ substitution yields
\begin{equation}
    \left[
    \int_{\mathbb{D}}
        |\mathcal{L}(\rho;L_c)|^2
        d^2\bsrho
    \right]^{1/2}
    =
    \sqrt{2\pi}
    N(L_c)
    L_c
    \left[
    \int_0^\infty
        |f(u)|^2
        u du
    \right]^{1/2}
    .
\end{equation}
Under the $N(L_c) \propto L_c^{-2}$ condition, we find that
the norm of $\mathcal{L}$ must be proportional to $L_c^{-1}$,
as in the case of the Gaussian correlation function, Eq.~(\ref{eq:Lc}).
Now, however, we see that this must hold for any correlation function that approaches a Dirac delta function in the $L_c\rightarrow 0$ limit.
We then conclude that any spatial correlation function with this behavior must decrease the near field TERS signal as $L_c$ increases, based on the intepretation of TERS as the inner product of $\mathcal{J}$ and $\mathcal{L}$ of Eq.~(\ref{eq:detected_signal_rho}).
This however does not mean that any arbitrary spatial correlation function leads to the same conclusion, since the delta function approach condition might not be necessary,
as with the Lorentzian correlation.

The appropriate choice of correlation function may depend on each experimental setting,
because being a correlation function of the polarizability,
it depends not only on the phonon propagation coherence, but also on the spatial coherence of the excitation field.
Therefore further work in this direction is needed to investigate whether to pick a Gaussian-like or a Lorentzian-like correlation function, or even some convolution of both.

\section{Conclusion}\label{sec:conclusion}

We have presented an extension of the TERS theoretical model to include the out-of-plane Raman response of 2D materials,
a necessary step if the model is applied to TERS measurements on TMDs.
On top of that, we could write the TERS near field scattered spectral density signal as an inner product between two functions, one having propagation and mode symmetry information, which could be written in exact analytical form, and the other having polarizability coherence length information.
The most important advantages of this interpretation are two.
One is that it is easy to calculate numerically, because it departs from analytical expressions.
The other is that it is now possible to easily experiment with different polarizability correlation functions.
We have shown that the model is sensitive to the choice of the correlation function, and this choice may influence how $L_c$ changes the TERS measurements.
So it is crucial to proceed with a deeper theoretical investigation of these functions,
and having a cheap computational tool to test them will be essential.
Our model also points a way to investigate the valid correlation functions experimentally, as an isolated manipulation of $L_c$, for instance with a manipulation of the doping in the sample~\cite{nadas2023}, could sort this question out.

An extensive analysis of how the physical parameters in the model change the predictions for experimental measurements was also made.
This analysis, summarized in Fig.~\ref{fig:params} and \ref{fig:enhancement}, can serve as a useful guide in the fitting of TERS experimental data,
as we did in the fittings of Fig.~\ref{fig:publionadas}.
Because the model is too complex with a very large number of physical parameters,
having an intuition of how changes in the parameters change the model output is essential to avoid unrealistic fittings.
It also shows, importantly,
that a strong TERS enhancement is a necessary condition for investigating the physics of a TERS sample,
which means surveying the difference in TERS signals, here represented by differences in tip approach curves.
Whether the enhancement is due to a large factor $f_e$,
a large tip radius,
a large relative Raman response $\nu$,
or bringing the tip really close to the sample,
the stronger the enhancement, the better one is able to distinguish the effect of a physical parameter on different Raman peaks,
for instance the coherence length $L_c$.

In particular, we show that the coherence length $L_c$ is the most important factor to get a large separation between normalized tip approach TERS curves of A$_1$ and E$_{2g}$ modes,
an effect of the destructive interference in the signal of the E$_{2g}$ mode~\cite{cancado2014},
consistent with the results of Ref.~\cite{ribeironeto2019}.
This means that any data in which the curves have a large separation will demand $L_c>0$, while closer curves might be fitted with $L_c=0$,
which reinforces the importance of the normalized TERS curves comparison as a testifier of $L_c>0$ in the sample~\cite{beams2014,cancado2014},
especially in graphene, in which $\nu\approx 0$.

It also comes out from our analysis that a large out-of-plane Raman response $\nu$ in the sample implies a large enhancement ratio between modes A$_1$ and E$_{2g}$,
since only mode A$_1$ depends on $\nu$ and thus a large $\nu$ increases the difference in the enhancements.
In addition, the most important factor to increase the enhancement ratio is the tip enhancement factor $f_e$, also consistent with Ref.~\cite{ribeironeto2019}, because it dramatically increases the TST term, which increases more for the A$_1$ mode,
and if it is associated with $\nu\geq 1$ it leads to very high enhancement ratios.
In turn, an increase in $L_c$ drives the enhancement ratio towards unity, which is a feature of the model predicting a decrease in the absolute TERS signal of both modes as $L_c$ increases.

A comparison with experimental data shows that the model is sound and can fit them with reasonable physical parameters.
However, there are still aspects that must be investigated, as in the data from Fig.~\ref{fig:publionadas}, in which graphene measurements from Ref.~\cite{publio2022} demand a coherence length $L_c>0$, while those from Ref.~\cite{nadas2025} can be fit with $L_c=0$.
The origin of $L_c$ is still elusive, whether it stems from the isolated phonon dynamics or from a coupled dynamics induced by the electric field, and the present model will be useful to address future measurements, possibly on TMDs, to compare with the previous literature \cite{publio2022,nadas2025}.

\section*{Acknowledgments}
The authors acknowledge Denis Basko for helpful discussions.
This work was supported by Conselho Nacional de Desenvolvimento Cientíﬁco e Tecnológico (CNPq, Grants No. 408697/2022-9, No. 421469/2023-4, and No. 308445/2023-6),
and Fundação de Amparo à Pesquisa do Estado de Minas Gerais (FAPEMIG, Grants No. APQ-04852-23, and No. RED-00081-23).

\section*{Data availability statement}
The data appearing in Fig.~\ref{fig:publionadas}, collected in Ref.~\cite{publio2022,nadas2025}, are not publicly available upon publication because it is not technically feasible and/or the cost of preparing, depositing, and hosting the data would be prohibitive within the terms of this research project. The data are available from the authors upon reasonable request.

\appendix

\section{Focused radially polarized Gaussian beam}\label{app:beam}

In TERS experiments, a radially polarized Gaussian beam is focused on the tip in order to obtain a large electric field component along the tip axis.
The tip is especially sensitive to this component, which results in a large field being scattered from the tip to the sample in the TST and ST terms.
In the TS term, the sample responds directly to the incident field, so if the sample Raman tensor has a $z$ component, it might contribute significantly to the TERS signal.
We provide, in this appendix, a justification for why the component of the focused field parallel to the $xy$ sample plane can be ignored in our treatment.

A radially polarized Gaussian doughnut mode on plane $xy$ propagating in the $z$ direction, after focusing, is \cite{novotny}
\begin{equation}\label{eq:field_radial}
    \mathbf{E} (r, \phi_r, z)
    =
    -\frac{ik f^2}{2 c w_0} \sqrt{\frac{n_1}{n_2}} E_0 e^{-ik f}
    \begin{bmatrix}
        i I_r (r,z) \cos\phi_r\\
        i I_r (r,z) \sin\phi_r\\
        - I_z (r,z) \\
    \end{bmatrix},
\end{equation}
where $(r,\phi_r)$ are radial coordinates at the $xy$ plane,
$k$ is the beam wave vector amplitude, $w_0$ is the Gaussian beam waist before focusing, $f$ is the objective lens focal distance, $n_1$ and $n_2$ are the refractive indices before and after the lens, respectively, and
\begin{subequations}\label{eq:field_components}
\begin{eqnarray}
    I_r (r, z)
    =&&
    4 \int\limits_0^{\theta_{\T{max}}}
    f_w(\theta) (\cos\theta)^{-1/2} (\sin\theta)^2
    \notag\\&&\times
    \cos\theta J_1(kr\sin\theta)
    e^{ikz\cos\theta}
    d\theta,
\end{eqnarray}
\begin{eqnarray}
    I_z (r, z)
    =&&
    4 \int\limits_0^{\theta_{\T{max}}}
    f_w(\theta) (\cos\theta)^{-1/2} (\sin\theta)^2
    \notag\\&&\times
    \sin\theta J_0(kr\sin\theta)
    e^{ikz\cos\theta}
    d\theta,
\end{eqnarray}
\end{subequations}
where $\theta_{\T{max}}$ is the largest polar angle captured by the lens (related to the NA by NA = $n_2 \sin\theta_{\T{max}}$), and
\begin{subequations}
\begin{equation}
    f_w(\theta)
    =
    e^{-\frac{1}{f_0^2} \frac{\sin^2\theta}{\sin^2\theta_{\T{max}}}},
\end{equation}
\begin{equation}
    f_0 = \frac{w_0}{f\sin\theta_{\T{max}}}.
\end{equation}
\end{subequations}

\begin{figure*}[!ht]
\begin{center}
    \includegraphics[width=0.9\textwidth]{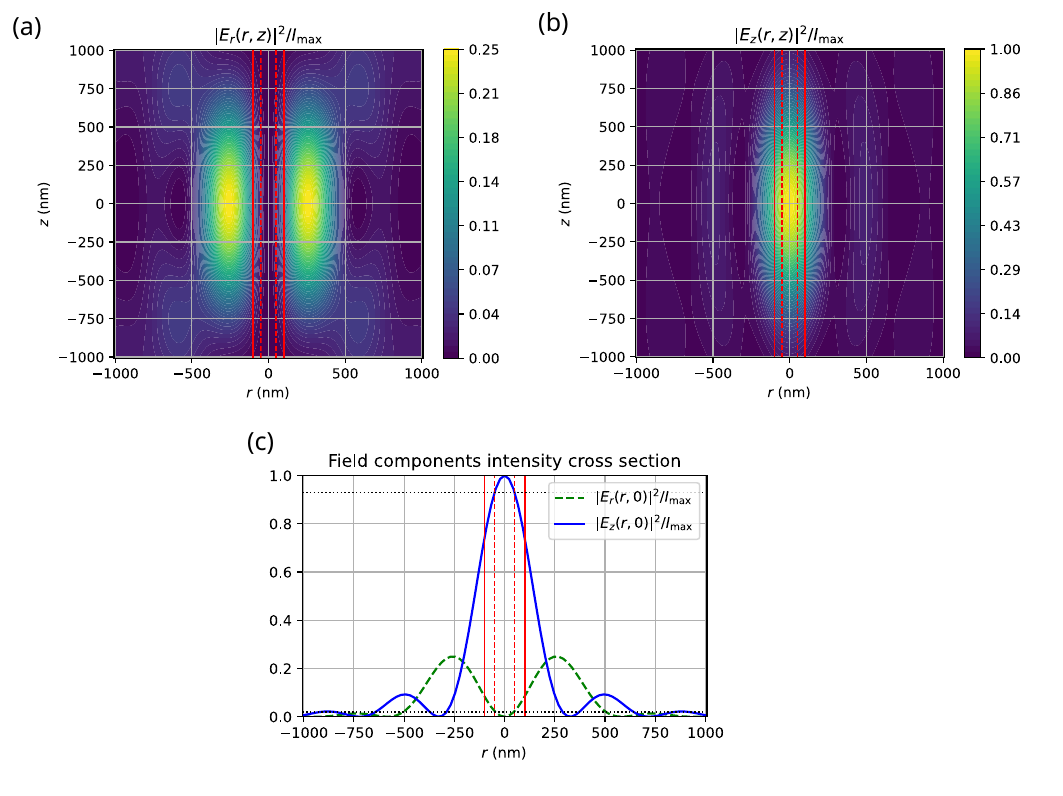}
\end{center}
    \caption{\label{fig:field_rz_plane}
        Intensity of components of the focused radially polarized Gaussian beam, all quantities normalized by the maximum field intensity, $I_{\T{max}} \equiv |\mathbf{E}(0,0)|^2$.
        In (a) we plot the radial component intensity and in (b) the $z$ component intensity---note the difference in the color scale between the two plots.
        In (c), we plot a cross section of the field components intensities at $z=0$.
        The vertical red lines mark a 50 nm (dashed line) and a 100 nm (solid line) radius around the beam axis, where the tip should be located.
        The horizontal black dotted lines mark the normalized intensities of 0.02 (bottom) and 0.93 (top), marking the respective values of the $r$ and $z$ components at the 50 nm radius.
    }
\end{figure*}

In Fig.~\ref{fig:field_rz_plane}~(a) and (b) we plot the intensity of the parallel and perpendicular components of the field in the $rz$ plane, calculated numerically from Eq.~\ref{eq:field_components}, normalized by the maximum intensity of the perpendicular field.
In Fig.~\ref{fig:field_rz_plane}~(c) we plot the intensity of the perpendicular ($z$, blue solid line) and radial ($r$ coordinate, or $xy$ plane, green dashed line) field components at the beam waist, $z=0$, normalized by the highest possible intensity.
The plot parameters that we use are a laser wavelength of $\lambda = $ 632.8 nm, yielding $k = 9.93 \times 10^{-3}$ rad/nm,
laser beam waist width $w_0 = 10$ mm,
objective lens focal distance $f = $ 170 $\mu$m,
objective NA 1.4,
and refractive index of medium after the lens (same as glass) $n = 1.52$.
These parameters result in a maximum lens angle $\theta_{\T{max}} = 33.5^\circ$, or $\sin\theta_{\T{max}} = 0.92$, from NA $ = n\sin\theta_{\T{max}}$, and $f_0 = 64$.
Note that \ref{eq:field_components} is a function of $kr$,
so Fig.~\ref{fig:field_rz_plane} can be easily adapted to other wavelengths by rescaling the axes.

The tip and sample are placed as close as possible to the beam waist at $z=0$, where the field is strongest, and a shift of tens of nanometers above or below the $z$ axis should not alter much this picture, as seen in Fig.~\ref{fig:field_rz_plane}~(a) and (b).
The vertical red lines mark a 50 nm (dashed line) and a 100 nm (solid line) radius around the beam axis.
If a 10 nm radius tip is well placed at the beam axis and aligned with it, it should be localized within the 50 nm radius and respond to the full $E_z$ component.
The horizontal black dotted lines in Fig.~\ref{fig:field_rz_plane}~(c) mark the normalized intensities of 0.02 and 0.93,
marking the respective values of the $r$ and $z$ components at the 50 nm radius,
meaning that, within this radius around the beam axis, $|E_r|^2 \lesssim 0.022 |E_z|^2$.
Even within a 100 nm radius, one has $|E_r|^2 \lesssim 0.13 |E_z|^2$,
so $E_r$ can be quite safely neglected in comparison with $E_z$.
In Ref. \cite{publio2022}, it was considered that $|E_r|^2 / |E_z|^2 \approx 1/3$, but we have shown that this ratio is at least 10 times smaller around the tip, where the near field is significant, so in the present article we only consider the $z$ component of the incident field.

\section{Details of the calculation of TST, ST, and TS propagation functions}\label{app:F_terms}

In this Appendix, we provide the detailed calculations of the $F$ propagation functions for the TST, ST, and TS contributions,
according to Eq.~(\ref{eq:F_contributions})
and the Green functions of Eq.~(\ref{eq:GNF_text}) and (\ref{eq:GFF}).

To take advantage of the cylindrical symmetry of the system,
we will use adimensional polar coordinates in the plane,
$\varrho \equiv \rho/z$,
and $\varphi_\varrho$ defined by $r_i/z \equiv [\delta_{iz} + (1-\delta_{iz})\varrho] \psi_i(\varphi_\varrho)$,
where $\psi_x(\varphi_\varrho) \equiv \cos(\varphi_\varrho)$,
$\psi_y(\varphi_\varrho) \equiv \sin(\varphi_\varrho)$
and $\psi_z(\varphi_\varrho) \equiv 1$.
In this notation, which will be used in Appendixes~\ref{app:F_terms}--\ref{app:bessel}, while $\delta_{iz}$ survives for $i=z$,
$(1-\delta_{iz})$ identifies the term that survives when $i\neq z$.
Then, Eq.~(\ref{eq:GNF_text}) can be rewritten as
\begin{eqnarray}\label{eq:GNF}
    G_{ij}^{\text{NF}} (\varrho, \varphi_\rho, z; \omega)
    = \frac{c^2}{4\pi n^2\omega^2}
    \left( \frac{1}{z^3} \right)
    \Bigg\{
        &-& \delta_{ij} \frac{1}{( \varrho^2 + 1 )^{3/2}}
        \notag\\
        &+&
        3
        [\delta_{iz} + (1-\delta_{iz})\varrho]
        [\delta_{jz} + (1-\delta_{jz})\varrho]
        \frac{
            \psi_i(\varphi_\varrho) \psi_j(\varphi_\varrho)
        }{( \varrho^2 + 1 )^{5/2}}
    \Bigg\},\notag\\
\end{eqnarray}
which seems cumbersome, but will allow us to solve some integrals.

\subsection{Tip-sample-tip (TST) term}\label{sec:TST}

We take the spatial Fourier transform of Eq.~(\ref{eq:FTST}),
\begin{eqnarray}
    F_{ijk}^{\text{TST}}(\mb{q}; \mb{r}_0, \omega_s, \omega)
    &=&
    \frac{\omega^2 \omega_s^2}{\varepsilon_0^2 c^4}
    \left( \alpha^{\text{tip}}_{zz} \right)^2
    G^{\text{FF}}_{kz}(\mbr_0,\mb{0};\omega_s)
    E_z (\omega)
    \\
    && \times
    \int_\mathbb{D}
        G^{\text{NF}}_{zj}(z\mbh{z},\bsrho;\omega_s)
        G^{\text{NF}}_{iz}(\bsrho,z\mbh{z};\omega)
    e^{-i \mb{q}\cdot \bsrho }
    d^2 \bsrho,
\end{eqnarray}
where the far field does not vary much so we can take it as coming from the origin $\mb{0}$.
We define the adimensional cylindrical coordinates in the wave vector space as $\xi \equiv |\mb{q}|z$,
and $\varphi_\xi$ defined by $q_iz \equiv \xi \psi_i(\varphi_\xi)$,
with $i \in \{x,y\}$.
Plugging the propagators from Eq.~(\ref{eq:GNF}),
\begin{equation}\label{eq:FTST_h}
    F_{ijk}^{\text{TST}}(\mb{q}; \mb{r}_0, \omega_s, \omega)
    =
    \left(
        \frac{\alpha^{\text{tip}}_{zz}}
        {4\pi \varepsilon}
    \right)^2
    \delta_{kz} g^{\text{FF}}_{z}
    E_z (\omega)
    \left(\frac{\pi}{z^4}\right)
    h_{ij}(\xi,\varphi_\xi),
\end{equation}
where we used $\varepsilon \equiv n^2\varepsilon_0$,
and defined
\begin{eqnarray}
    h_{ij}(\xi,\varphi_\xi) \equiv
    \frac{1}{\pi} \int_0^\infty \int_0^{2\pi}
    \Bigg\{
        &9& [\delta_{iz} + (1-\delta_{iz})\varrho]
            [\delta_{jz} + (1-\delta_{jz})\varrho]
            \frac{
                \psi_i(\varphi_\varrho) \psi_j(\varphi_\varrho)
            }{( \varrho^2 + 1 )^5}
        + \delta_{iz} \delta_{jz}
            \frac{1}{( \varrho^2 + 1 )^3}
        \notag\\
        &-&
        3 \delta_{iz} [\delta_{jz} + (1-\delta_{jz})\varrho]
        \frac{
            \psi_j(\varphi_\varrho)
        }{( \varrho^2 + 1 )^4}
        - 3 [\delta_{iz} + (1-\delta_{iz})\varrho] \delta_{jz}
        \frac{
            \psi_i(\varphi_\varrho)
        }{( \varrho^2 + 1 )^4}
    \Bigg\}\notag\\
        &\times& e^{-i \xi\rho \cos(\varphi_\xi-\varphi_\rho) }
    \varrho d\varrho d\varphi_\varrho.
\end{eqnarray}
We can write this expression in a more insightful manner, to make it easier to identify the contributions of each component,
\begin{eqnarray}\label{eq:h_def}
    h_{ij}(\xi, \varphi_\xi)
    =
    \frac{1}{\pi} \int_0^\infty \int_0^{2\pi}
    \Bigg\{
        &\delta_{iz}& \delta_{jz}
            \left[
                9 \frac{1}{( \varrho^2 + 1 )^5}
                +
                \frac{1}{( \varrho^2 + 1 )^3}
                - 6
                \frac{1}{( \varrho^2 + 1 )^4}
            \right]
            \notag\\
        &+&
            \left[ (1-\delta_{iz}) \delta_{jz} \psi_i(\varphi_\varrho)
                + \delta_{iz} (1-\delta_{jz}) \psi_j(\varphi_\varrho)
            \right]
            \left[
                9 \frac{\varrho}{( \varrho^2 + 1 )^5}
                - 3 \frac{\varrho}{( \varrho^2 + 1 )^4}
            \right]
            \notag\\
        &+& 9 (1-\delta_{iz}) (1-\delta_{jz})
            \psi_i(\varphi_\varrho) \psi_j(\varphi_\varrho)
            \frac{\rho^2}{( \varrho^2 + 1 )^5}
    \Bigg\}
    e^{-i \xi\rho \cos(\phi_\xi-\phi_\rho) }
    \varrho d\varrho d\phi_\varrho.\notag\\
\end{eqnarray}
The top term between the curly brackets in Eq.~(\ref{eq:h_def}) is the one for $i=j=z$, the middle one for $i \in \{x,y\}$ and $j=z$ and vice versa,
and the bottom one for $i,j \in \{x,y\}$.
If we set $i,j \in \{x,y\}$, we get back to Ref. \cite{cancado2014,publio2022}, where the sample does not respond in direction $z$.

To solve these integrals analytically, we use the Jacobi-Anger expansion \cite{cuyt},
\begin{equation}\label{eq:jacobi}
    e^{-i \xi\varrho \cos(\varphi_\xi -\varphi_\varrho)}
    = J_0(\xi\varrho) + 2\sum\limits_{n=1}^{\infty} (-i)^{n} J_n(\xi\varrho) \cos( n \varphi_\xi - n\varphi_\varrho),
\end{equation}
so that we can solve the angular integral separately from the radial one.
For that matter, we must use trigonometric identities and orthogonality relations for trigonometric functions (see Appendix \ref{app:angular}), arriving at
\begin{eqnarray}
    h_{ij}(\xi, \varphi_\xi)
    =
    \int_0^\infty
    \Bigg\{
        &\delta_{iz}& \delta_{jz}
            \left[ 2 J_0(\xi\varrho) \right]
            \left[
                9 \frac{1}{( \varrho^2 + 1 )^5}
                +
                \frac{1}{( \varrho^2 + 1 )^3}
                - 6
                \frac{1}{( \varrho^2 + 1 )^4}
            \right]
            \notag\\
        &+&
            \left[ (1-\delta_{iz}) \delta_{jz} \psi_i(\varphi_\xi)
                + \delta_{iz} (1-\delta_{jz}) \psi_j(\varphi_\xi)
            \right]
            \left[ - 2i \right]
            \left[
                9 \frac{\varrho}{( \varrho^2 + 1 )^5}
                - 3 \frac{\varrho}{( \varrho^2 + 1 )^4}
            \right]
            \notag\\
        &+& 9 (1-\delta_{iz}) (1-\delta_{jz})
            \left[
                \delta_{ij} J_0(\xi\varrho)
                - \tau_{ij}(\varphi_\xi) J_2(\xi\varrho)
            \right]
            \frac{\rho^2}{( \varrho^2 + 1 )^5}
    \Bigg\}
    \varrho d\varrho,\notag\\
\end{eqnarray}
where
\begin{equation}\label{eq:tau_def}
   \tau_{ij}(\phi_\xi)
    \equiv
    \begin{bmatrix}
        \cos(2\phi_\xi) & \sin(2\phi_\xi) \\
        \sin(2\phi_\xi) & -\cos(2\phi_\xi) \\
    \end{bmatrix}_{ij}.
\end{equation}

In Appendix \ref{app:bessel}, we show how the solutions of these integrals come from the theory of Hankel (or Bessel) transforms \cite{piessens_hankel, davies_transforms, relton_bessel}.
With that we can write an exact analytical expression for $h_{ij}(\xi, \varphi_\xi)$,
\begin{eqnarray}\label{eq:h_analytical}
    h_{ij}(\xi, \varphi_\xi)
    =
    &2&
        \delta_{iz} \delta_{jz}
        \left[
            \frac{9}{384} \xi^4 K_4(\xi)
            - \frac{6}{48} \xi^3 K_3(\xi)
            + \frac{1}{8} \xi^2 K_2(\xi)
        \right]
        \notag\\
    &-&
        2i
        \left[ (1-\delta_{iz}) \delta_{jz} \psi_i(\varphi_\xi)
            + \delta_{iz} (1-\delta_{jz}) \psi_j(\varphi_\xi)
        \right]
        \left[
            \frac{9}{384} \xi^4 K_3(\xi)
            - \frac{3}{48} \xi^3 K_2(\xi)
        \right]
        \notag\\
    &+& (1-\delta_{iz}) (1-\delta_{jz})
        \left[
            \delta_{ij}
            \frac{9}{384} \left( 2 \xi^3 K_3(\xi) -\xi^4 K_2(\xi) \right)
            - \tau_{ij}(\varphi_\xi)
            \frac{9}{384} \xi^4 K_2(\xi)
        \right]
    ,\notag\\
\end{eqnarray}
where $K_\nu(\xi)$ is the modified Bessel function of the second kind of order $\nu$.

\subsection{Tip-sample (ST) term}\label{sec:ST}

To write the ST term, we take the spatial Fourier transform of Eq.~(\ref{eq:FST}),
\begin{equation}
    F_{ijk}^{\text{ST}} (\mb{q}; \mbr_0, \omega_s,\omega)
    =
    \frac{\omega^2}{\varepsilon_0 c^2} \alpha^{\text{tip}}_{zz}
    G^{\text{FF}}_{kj}(\mbr_0,\mb{0};\omega_s)
    E_z (\omega)
    \int_\mathbb{D}
        G^{\text{NF}}_{iz}(\bsrho,z\mbh{z};\omega)
    e^{-i \mb{q}\cdot \bsrho }
    d^2 \bsrho,
\end{equation}
where again we assume the far field as coming from the origin $\mb{0}$.
Plugging the propagator from Eq.~(\ref{eq:GNF}),
\begin{equation}\label{eq:FST_a}
    F_{ijk}^{\text{ST}} (\mb{q}; \mbr_0, \omega_s,\omega)
    =
    \left( \frac{\alpha^{\text{tip}}_{zz}}{4\pi\varepsilon} \right)
    \delta_{kj} [ g^{\text{FF}}_z \delta_{jz} + g^{\text{FF}}_{\parallel} (1-\delta_{jz})]
    E_z (\omega)
    \left( \frac{\pi}{z} \right)
    a_i(\xi,\varphi_\xi),
\end{equation}
where we defined
\begin{eqnarray}\label{eq:a_def}
    a_i(\xi,\varphi_\xi)
    \equiv
    \frac{1}{\pi} \int_0^\infty \int_0^{2\pi}
    \Bigg\{
        &\delta_{iz}&
        \left[
            3 \frac{1}{( \varrho^2 + 1 )^{5/2}}
            - \frac{1}{( \varrho^2 + 1 )^{3/2}}
        \right]
        \notag\\
        &+& 3(1-\delta_{iz})
        \psi_i(\varphi_\varrho)
        \frac{ \varrho }{( \varrho^2 + 1 )^{5/2}}
    \Bigg\}
    e^{-i \xi\rho \cos(\varphi_\xi-\varphi_\rho) }
    \varrho d\varrho d\varphi_\varrho.
\end{eqnarray}
In Eq.~(\ref{eq:a_def}), if we set $i \in \{x,y\}$, only the last term between the curly brackets survives and we get back to Ref. \cite{cancado2014,publio2022}, where the sample $z$ component does not respond.

Using the results from Appendix \ref{app:angular}, the angular integrals can be solved analytically,
\begin{eqnarray}
    a_i(\xi,\varphi_\xi)
    =
    \int_0^\infty
    \Bigg\{
        &\delta_{iz}&
        \left[ 2J_0(\xi\varrho) \right]
        \left[
            3 \frac{1}{( \varrho^2 + 1 )^{5/2}}
            - \frac{1}{( \varrho^2 + 1 )^{3/2}}
        \right]
        \notag\\
        &-& 6i (1-\delta_{iz})
        \psi_i(\varphi_\xi)
        J_1(\xi\varrho)
        \frac{ \varrho }{( \varrho^2 + 1 )^{5/2}}
    \Bigg\}
    \varrho d\varrho d\varphi_\varrho.
\end{eqnarray}
The results from Appendix \ref{app:bessel} then provide the exact analytical expression for $a_i(\xi,\varphi_\xi)$,
\begin{equation}\label{eq:a_analytical}
    a_i(\xi,\varphi_\xi)
    =
    2
    \left[
        \delta_{iz}
        - i (1-\delta_{iz})
        \psi_i(\varphi_\xi)
    \right]
    \xi e^{-\xi}
    .
\end{equation}

\subsection{Sample-tip (TS) term}\label{sec:TS}

Taking the Fourier transform of Eq.~(\ref{eq:FTS}),
\begin{equation}
    F_{ijk}^{\text{TS}} (\mb{q}; \mbr_0, \omega_s,\omega)
    =
    \frac{\omega_s^2}{\varepsilon_0 c^2} \alpha^{\text{tip}}_{zz}
    G^{\text{FF}}_{kz}(\mbr_0,\mb{0};\omega_s)
    \int_\mathbb{D}
        G^{\text{NF}}_{zj}(z\mbh{z},\bsrho;\omega_s)
        E_i (\bsrho, \omega)
        e^{-i \mb{q}\cdot \bsrho }
    d^2 \bsrho,
\end{equation}
we see, distinctly from the TST and ST terms, that the Field may vary with $\bsrho$, because here it reaches the sample first.
However, we know that the $xy$-plane field components are very weak
in the vicinity of the tip, where the near field propagator contributes with relevant magnitude,
so we will work only with the $z$ component, which we may consider as constant over a 50 nm radius around the tip (see Appendix \ref{app:beam}).
At this distance the near field contribution certainly vanishes, so only atoms very close to the tip will contribute to TS.
Then, plugging the propagator from Eq.~(\ref{eq:GNF}),
\begin{equation}\label{eq:FTS_a}
    F_{ijk}^{\text{TS}} (\mb{q}; \mbr_0, \omega_s,\omega)
    =
    \left( \frac{\alpha^{\text{tip}}_{zz}}{4\pi\varepsilon} \right)
    \delta_{kz} g^{\text{FF}}_z
    \delta_{iz} E_z (\omega)
    \left( \frac{\pi}{z} \right)
    a_j(\xi,\varphi_\xi),
\end{equation}
with $a_j(\xi,\varphi_\xi)$ defined in Eq.~(\ref{eq:a_def}) and evaluated to Eq.~(\ref{eq:a_analytical}).

\section{Angular integrals}\label{app:angular}

We need to calculate three kinds of angular integrals,
\begin{subequations}
\begin{equation}\label{eq:Iij}
    I_{ij} =
    \int_0^{2\pi}
    \psi_i(\varphi_\varrho) \psi_j(\varphi_\varrho)
    e^{-i \xi\varrho \cos(\varphi_\xi -\varphi_\varrho)}
    d\varphi_\varrho,
\end{equation}
\begin{equation}\label{eq:Iiz}
    I_{iz} =
    \int_0^{2\pi}
    \psi_i(\varphi_\varrho)
    e^{-i \xi\varrho \cos(\varphi_\xi -\varphi_\varrho)}
    d\varphi_\varrho,
\end{equation}
\begin{equation}\label{eq:Izz}
    I_{zz} =
    \int_0^{2\pi}
    e^{-i \xi\varrho \cos(\varphi_\xi -\varphi_\varrho)}
    d\varphi_\varrho,
\end{equation}
\end{subequations}
with $i,j\in\{x,y\}$, $\psi_x(\varphi_\varrho) = \cos(\varphi_\varrho)$,
and $\psi_y(\varphi_\varrho) = \sin(\varphi_\varrho)$.
We start with the most complicated one, Eq.~(\ref{eq:Iij}).
Expansion of the exponential with the Jacobi-Anger expansion \cite{cuyt} [see Eq.~(\ref{eq:jacobi})] yields
\begin{eqnarray}
    I_{ij} &=&
    \int_0^{2\pi}
        \psi_i(\varphi_\varrho) \psi_j(\varphi_\varrho)
        J_0(\xi\varrho)
    d\varphi_\varrho
    \notag\\
    && +
    2 \sum_{n=1}^\infty
    (-i)^n
    \int_0^{2\pi}
        \psi_i(\varphi_\varrho) \psi_j(\varphi_\varrho)
        \cos( n \varphi_\xi - n \varphi_\varrho)
        J_n(\xi\varrho)
    d\varphi_\varrho
    .
\end{eqnarray}
We have three possible situations,
$\psi_i(\varphi_\varrho) \psi_j(\varphi_\varrho) = \cos^2(\varphi_\varrho) = [1 + \cos(2\varphi_\varrho)]/2$;
$\psi_i(\varphi_\varrho) \psi_j(\varphi_\varrho) = \sin^2(\varphi_\varrho) = [1 - \cos(2\varphi_\varrho)]/2$;
or
$\psi_i(\varphi_\varrho) \psi_j(\varphi_\varrho) = \sin(\varphi_\varrho) \cos(\varphi_\varrho) = \sin(2\varphi_\varrho)/2$.
For compactness, we write $\mathcal{C}_{n\kappa} \equiv \cos(n\varphi_\kappa)$ and $\mathcal{S}_{n\kappa} \equiv \sin(n\varphi_\kappa)$, with $\kappa = \{\xi, \varrho\}$,
and use the inner product definition $\langle f,g \rangle \equiv \int_0^{2\pi} f(\varphi_\varrho) g(\varphi_\varrho) d\varphi_\varrho$ to write
\begin{eqnarray}
    \int_0^{2\pi}
        \mathcal{C}^2(\varphi_\varrho)
        \mathcal{C}(n\varphi_\xi -n\varphi_\varrho)
    d\varphi_\varrho
    &=&
    \langle (1+\mathcal{C}_{2\varrho})/2, (\mathcal{C}_{n\xi}\mathcal{C}_{n\varrho} + \mathcal{S}_{n\xi}\mathcal{S}_{n\varrho}) \rangle
    \notag\\
    &=&
    \mathcal{C}_{n\xi}
    \left(
        \langle 1, \mathcal{C}_{n\varrho} \rangle
        + \langle \mathcal{C}_{2\varrho}, \mathcal{C}_{n\varrho} \rangle
    \right)/2
    +
    \mathcal{S}_{n\xi}
    \left(
        \langle 1, \mathcal{S}_{n\varrho} \rangle
        + \langle \mathcal{C}_{2\varrho}, \mathcal{S}_{n\varrho} \rangle
    \right)/2,
    \notag\\
\end{eqnarray}
and analogously for the other possibilities.

Using the orthogonality relations of cosine and sine functions, $\langle \mathcal{C}_{m\rho}, \mathcal{C}_{n\rho} \rangle = \pi \delta_{nm}$,
$\langle \mathcal{S}_{m\rho}, \mathcal{S}_{n\rho} \rangle = \pi \delta_{nm}$,
and
$\langle \mathcal{S}_{m\rho}, \mathcal{C}_{n\rho} \rangle = 0$,
for $m,n \geq 1$,
we find
\begin{subequations}
\begin{equation}
    \int_0^{2\pi} d{\varphi_\varrho}
    \cos^2(\varphi_\varrho)
    \cos(n\varphi_\xi -n\varphi_\varrho)
    =
    \pi \delta_{n0}
    + \frac{\pi}{2} \cos(2\varphi_\xi) \delta_{n2},
\end{equation}
\begin{equation}
    \int_0^{2\pi} d{\varphi_\varrho}
    \sin^2(\varphi_\varrho)
    \cos(n\varphi_\xi -n\varphi_\varrho)
    =
    \pi \delta_{n0}
    - \frac{\pi}{2} \cos(2\varphi_\xi) \delta_{n2},
\end{equation}
\begin{equation}
    \int_0^{2\pi} d{\varphi_\varrho}
    \cos(\varphi_\varrho)\sin(\varphi_\varrho)
    \cos(n\varphi_\xi -n\varphi_\varrho)
    =
    \frac{\pi}{2} \sin(2\varphi_\xi) \delta_{n2}.
\end{equation}
\end{subequations}
Integral $I_{ij}$ of Eq.~(\ref{eq:Iij}) is then evaluated to
\begin{equation}\label{eq:Iij_solution}
    I_{ij}
    =
    \pi
    \left[
        \delta_{ij} J_0(\xi\rho)
        -\tau_{ij}(\phi_\xi) J_2(\xi\rho)
    \right],
\end{equation}
with
\begin{equation}
   \tau_{ij}(\phi_\xi)
    \equiv
    \begin{bmatrix}
        \cos(2\phi_\xi) & \sin(2\phi_\xi) \\
        \sin(2\phi_\xi) & -\cos(2\phi_\xi) \\
    \end{bmatrix}_{ij}.
\end{equation}

The integral of Eq.~(\ref{eq:Iiz}) is simpler,
\begin{eqnarray}
    I_{iz} &=&
    \int_0^{2\pi}
        \psi_i(\varphi_\varrho)
        J_0(\xi\varrho)
    d\varphi_\varrho
    \notag\\
    && +
    2 \sum_{n=1}^\infty
    (-i)^n
    \int_0^{2\pi}
        \psi_i(\varphi_\varrho)
        \cos( n \varphi_\xi - n \varphi_\varrho)
        J_n(\xi\varrho)
    d\varphi_\varrho
    .
\end{eqnarray}
In this case, the integral for $n=0$ is null, and we need to check only $n \geq 1$,
\begin{eqnarray}
    \int_0^{2\pi} d{\varphi_\varrho}
    \psi_i(\varphi_\varrho)
    \cos( n\varphi_\xi - n\varphi_\varrho)
    &=&
    \mathcal{C}_{n\xi} \langle \psi_i, \mathcal{C}_{n\varrho} \rangle
    + \mathcal{S}_{n\xi} \langle \psi_i, \mathcal{S}_{n\varrho} \rangle
    \notag\\
    &=&
    \pi \psi_i(\varphi_\xi) \delta_{n1}.
\end{eqnarray}
The integral is then evaluated to
\begin{equation}\label{eq:Iiz_solution}
    I_{iz} =
    -2\pi i
    \psi_i(\phi_\xi)
    J_1(\xi\rho)
    .
\end{equation}

Finally, the integral of Eq.~(\ref{eq:Izz}) is the simplest,
\begin{eqnarray}
    I_{zz} &=&
    \int_0^{2\pi}
        J_0(\xi\varrho)
    d\varphi_\varrho
    \notag\\
    && +
    2 \sum_{n=1}^\infty
    (-i)^n
    \int_0^{2\pi}
        \cos( n \varphi_\xi - n \varphi_\varrho)
        J_n(\xi\varrho)
    d\varphi_\varrho
    ,
\end{eqnarray}
because all terms $n \geq 1$ yield zero, and
\begin{equation}\label{eq:Izz_solution}
    I_{zz}
    =
    2\pi J_0(\xi\varrho)
    .
\end{equation}

\section{Integrals of Bessel functions}\label{app:bessel}

In this Appendix, we provide exact analytical solutions to the integrals that appear Appendix~\ref{app:F_terms}.
In order to obtain the solutions of the TST terms, we will need the relation
\begin{equation}\label{eq:hankel_transform}
    \int_0^{\infty}
    \frac{\varrho^{\nu+1}}{(\varrho^2+a^2)^{\mu+1}}
    J_\nu(\xi\varrho)
    d\varrho
    =
    \frac{\xi^\mu a^{\nu-\mu}}{2^\mu \Gamma(\mu+1)}
    K_{\nu-\mu}(a\xi),
\end{equation}
where $K_\nu(\xi)$ is the modified Bessel function of the second kind, and which comes from the theory of Hankel (or Bessel) transforms
\cite{piessens_hankel}.
We will also use the transform of the solution of the Bessel equation \cite{davies_transforms},
\begin{equation}
    \int_0^{\infty}
    (-\varrho^2) f(\varrho)
    J_\nu(\xi\varrho)
    d\varrho
    =
    \frac{\partial^2 F_\nu(\xi)}{\partial\xi^2}
    +\frac{1}{\xi} \frac{\partial F_\nu(\xi)}{\partial\xi}
    -\frac{\nu^2}{\xi^2} F_\nu(\xi),
\end{equation}
where $F_\nu(\xi) \equiv \int_0^{\infty} f(\varrho) J_\nu(\xi\varrho) d\varrho$.

To calculate
\begin{equation}
    \int_0^{\infty}
    \frac{\varrho^2}{(\varrho^2+1)^{5}}
    J_0(\xi\varrho) \varrho d\varrho,
\end{equation}
we write it as
\begin{equation}
    -\int_0^{\infty}
    (-\varrho^2)
    \frac{\varrho}{(\varrho^2+1)^{5}}
    J_0(\xi\varrho) d\varrho.
\end{equation}
Then we can identify $\mu=4$ and $\nu=0$ to insert in Eq.~\eqref{eq:hankel_transform},
such that
\begin{equation}
    \int_0^{\infty}
    \frac{\varrho^2}{(\varrho^2+1)^{5}}
    J_0(\xi\varrho) \varrho d\varrho
    =
    -\frac{\partial^2 F^{\T{TST}}_{0}(\xi)}{\partial\xi^2}
    -\frac{1}{\xi} \frac{\partial F^{\T{TST}}_{0}(\xi)}{\partial\xi},
\end{equation}
with
\begin{equation}
    F^{\T{TST}}_{0}(\xi)
    =
    \int_0^{\infty}
    \frac{\varrho}{(\varrho^2+1)^{5}}
    J_0(\xi\varrho) d\varrho
    =
    \frac{\xi^4}{2^4 \Gamma(5)}
    K_{4}(\xi),
\end{equation}
where we used that $K_{-\nu}(\xi) = K_{\nu}(\xi)$ \cite{relton_bessel}.
Then, using
$\frac{d}{d\xi}(\xi^\nu K_\nu(\xi)) = -\xi^\nu K_{\nu-1}(\xi)$ \cite{relton_bessel},
one arrives at
\begin{equation}
    \int_0^{\infty}
    \frac{\varrho^2}{(\varrho^2+1)^{5}}
    J_0(\xi\varrho) \varrho d\varrho
    =
    \frac{1}{2^4 \Gamma(5)}
    \left[
        2\xi^3 K_3(\xi)
        -\xi^4 K_2(\xi)
    \right],
\end{equation}
in which ${2^4 \Gamma(5)} = 384$.

All other integrals are a direct application of Eq.~\eqref{eq:hankel_transform}, where we can use
\begin{subequations}
    \begin{equation}
        K_{1/2}(\xi) =
        \sqrt{\frac{\pi}{2\xi}}
        e^{-\xi},
    \end{equation}
    \begin{equation}
        K_{3/2}(\xi) =
        \sqrt{\frac{\pi}{2\xi}}
        \left( 1 + \frac{1}{\xi} \right) e^{-\xi},
    \end{equation}
\end{subequations}
in the ST ones.

\subsection{TST terms}

Below we list the exact analytical integral results needed in Appendix~\ref{app:F_terms} for the calculation of the TST terms.
\begin{equation}\label{eq:J0TST}
    \int_0^{\infty}
    \frac{\varrho^2}{(\varrho^2+1)^{5}}
    J_0(\xi\varrho) \varrho d\varrho
    =
    \frac{1}{384} \left[ 2 \xi^3 K_3(\xi) - \xi^4 K_2(\xi) \right]
\end{equation}
\begin{equation}
    \int_0^{\infty}
    \frac{\varrho^2}{(\varrho^2+1)^{5}}
    J_2(\xi\varrho) \varrho d\varrho
    =
    \frac{1}{384} \xi^4 K_2(\xi)
\end{equation}
\begin{equation}
    \int_0^{\infty}
    \frac{\varrho}{(\varrho^2+1)^{5}}
    J_1(\xi\varrho) \varrho d\varrho
    =
    \frac{1}{384} \xi^4 K_3(\xi)
\end{equation}
\begin{equation}
    \int_0^{\infty}
    \frac{\varrho}{(\varrho^2+1)^{4}}
    J_1(\xi\varrho) \varrho d\varrho
    =
    \frac{1}{48} \xi^3 K_2(\xi)
\end{equation}
\begin{equation}
    \int_0^{\infty}
    \frac{1}{(\varrho^2+1)^{5}}
    J_0(\xi\varrho) \varrho d\varrho
    =
    \frac{1}{384} \xi^4 K_4(\xi)
\end{equation}
\begin{equation}
    \int_0^{\infty}
    \frac{1}{(\varrho^2+1)^{3}}
    J_0(\xi\varrho) \varrho d\varrho
    =
    \frac{1}{8} \xi^2 K_2(\xi)
\end{equation}
\begin{equation}
    \int_0^{\infty}
    \frac{1}{(\varrho^2+1)^{4}}
    J_0(\xi\varrho) \varrho d\varrho
    =
    \frac{1}{48} \xi^3 K_3(\xi)
\end{equation}

\subsection{ST terms}
Below we list the exact analytical integral results needed in Appendix~\ref{app:F_terms} for the calculation of the ST terms.

\begin{equation}
    \int_0^{\infty}
    \frac{\varrho}{(\varrho^2+1)^{5/2}}
    J_1(\xi\varrho) \varrho d\varrho
    =
    \frac{1}{3} \xi e^{-\xi}
\end{equation}
\begin{equation}
    \int_0^{\infty}
    \frac{1}{(\varrho^2+1)^{5/2}}
    J_0(\xi\varrho) \varrho d\varrho
    =
    \frac{1}{3} (1+\xi) e^{-\xi}
\end{equation}
\begin{equation}
    \int_0^{\infty}
    \frac{1}{(\varrho^2+1)^{3/2}}
    J_0(\xi\varrho) \varrho d\varrho
    =
    e^{-\xi}
\end{equation}

\section{Derivation of the tensorial contribution to the TERS signal}\label{app:I_calculation}

The goal of this appendix is to derive a compact form for the $\mc{I}^{(\eta)}$ functions, defined in Eq.~(\ref{eq:mcI}), which carry all the tensorial properties of the TERS scattering of mode $\eta$.
First we must write the explicit form of the total $F_{ijk}$, substituting Eq.~(\ref{eq:FTST_h}), (\ref{eq:FST_a}) and (\ref{eq:FTS_a}) into Eq.~(\ref{eq:F_sum_of_contributions}),
\begin{eqnarray}\label{eq:F_total}
    F_{ijk} (\mb{q}; \mbr_0, \omega_s,\omega)
    &=&
    \delta_{kz}
    g^{\text{FF}}_{z}
    E_z (\omega)
    \notag\\
    &&\times \left\{
    \left(
        \frac{\alpha^{\text{tip}}_{zz}}
        {4\pi \varepsilon}
    \right)^2
    \frac{\pi}{z^4}
    h_{ij}(\xi,\varphi_\xi)
    + \left( \frac{\alpha^{\text{tip}}_{zz}}{4\pi\varepsilon} \right)
    \frac{\pi}{z}
    \left[
        \delta_{jz} a_i(\xi,\varphi_\xi)
        + a_j(\xi,\varphi_\xi) \delta_{iz}
    \right]
    \right\}
    \notag\\
    &&+
    (1-\delta_{kz}) g^{\text{FF}}_{\parallel}
    E_z (\omega)
    \left( \frac{\alpha^{\text{tip}}_{zz}}{4\pi\varepsilon} \right)
    \frac{\pi}{z}
    \delta_{kj} a_i(\xi,\varphi_\xi).
\end{eqnarray}

Using Eq.~(\ref{eq:F_total}) and the Raman tensors, Eq.~(\ref{eq:alpha_matrices}), in (\ref{eq:F_sums}), we can write
\begin{subequations}\label{eq:Falpha}
    \begin{eqnarray}\label{eq:Falpha_A1}
        \sum_{i,j} F_{ijk} \alpha_{ij}^{(A_1)}
        =
        \alphabar^{(A_1)} E_z (\omega) &\Bigg\{&
        \delta_{kz}
        g^{\text{FF}}_{z}
        \left(
            \frac{\alpha^{\text{tip}}_{zz}}
            {4\pi \varepsilon}
        \right)^2
        \frac{\pi}{z^4}
        \left[ h_{xx}(\xi,\varphi_\xi) + h_{yy}(\xi,\varphi_\xi) \right]
        \notag\\
        &&+
        g^{\text{FF}}_{\parallel}
        \left( \frac{\alpha^{\text{tip}}_{zz}}{4\pi\varepsilon} \right)
        \frac{\pi}{z}
        \left[ \delta_{kx} a_x(\xi,\varphi_\xi) + \delta_{ky} a_y(\xi,\varphi_\xi) \right]
        \notag\\
        &&+
        \nu
        \delta_{kz}
        g^{\text{FF}}_{z}
        \left[
            \left(
                \frac{\alpha^{\text{tip}}_{zz}}
                {4\pi \varepsilon}
            \right)^2
            \frac{\pi}{z^4}
            h_{zz}(\xi,\varphi_\xi)
            + 2
            \left( \frac{\alpha^{\text{tip}}_{zz}}{4\pi\varepsilon} \right)
            \frac{\pi}{z}
            a_z(\xi,\varphi_\xi)
        \right]
        \Bigg\},
        \notag\\
    \end{eqnarray}
    \begin{eqnarray}
        \sum_{i,j} F_{ijk} \alpha_{ij}^{(E_{2g},x)}
        =
        \alphabar^{(E_{2g})} E_z (\omega) &\Bigg\{&
        \delta_{kz}
        g^{\text{FF}}_{z}
        \left(
            \frac{\alpha^{\text{tip}}_{zz}}
            {4\pi \varepsilon}
        \right)^2
        \frac{\pi}{z^4}
        \left[ h_{xy}(\xi,\varphi_\xi) + h_{yx}(\xi,\varphi_\xi) \right]
        \notag\\
        &&+
        g^{\text{FF}}_{\parallel}
        \left( \frac{\alpha^{\text{tip}}_{zz}}{4\pi\varepsilon} \right)
        \frac{\pi}{z}
        \left[ \delta_{kx} a_y(\xi,\varphi_\xi)
            + \delta_{ky} a_x(\xi,\varphi_\xi) \right]
        \Bigg\},
    \end{eqnarray}
    \begin{eqnarray}
        \sum_{i,j} F_{ijk} \alpha_{ij}^{(E_{2g},y)}
        =
        \alphabar^{(E_{2g})} E_z (\omega) &\Bigg\{&
        \delta_{kz}
        g^{\text{FF}}_{z}
        \left(
            \frac{\alpha^{\text{tip}}_{zz}}
            {4\pi \varepsilon}
        \right)^2
        \frac{\pi}{z^4}
        \left[ h_{xx}(\xi,\varphi_\xi) - h_{yy}(\xi,\varphi_\xi) \right]
        \notag\\
        &&+
        g^{\text{FF}}_{\parallel}
        \left( \frac{\alpha^{\text{tip}}_{zz}}{4\pi\varepsilon} \right)
        \frac{\pi}{z}
        \left[ \delta_{kx} a_x(\xi,\varphi_\xi)
            - \delta_{ky} a_y(\xi,\varphi_\xi) \right]
        \Bigg\},
    \end{eqnarray}
\end{subequations}
where we left the $\nu$-dependent terms separated from the others, for easy identification.
One $\nu$-dependent term comes from a TST contribution, identified by the $h_{zz}$ function in Eq.~(\ref{eq:Falpha_A1}),
and the other comes from the sum of ST and TS contributions, which are equal in magnitude, identified by the factor 2 and the function $a_z$ in Eq.~(\ref{eq:Falpha_A1}).

Using Eq.~(\ref{eq:Falpha}) in Eq.~(\ref{eq:mcI}), we can write
\begin{subequations}\label{eq:Ieta_step1}
    \begin{eqnarray}\label{eq:IA1_step1}
        \mc{I}^{(A_1)}
        =
        |\alphabar^{(A_1)}|^2 |E_z (\omega)|^2
        &\Bigg\{&
        \left|g^{\text{FF}}_{z}\right|^2
        \Bigg|
        \left(
            \frac{\alpha^{\text{tip}}_{zz}}
            {4\pi \varepsilon}
        \right)^2
        \frac{\pi}{z^4}
        \left[ h_{xx}(\xi,\varphi_\xi) + h_{yy}(\xi,\varphi_\xi) + \nu h_{zz}(\xi,\varphi_\xi) \right]
        \notag\\
        &&+
        2 \nu
            \left( \frac{\alpha^{\text{tip}}_{zz}}{4\pi\varepsilon} \right)
            \frac{\pi}{z}
            a_z(\xi,\varphi_\xi)
        \Bigg|^2
        \notag\\
        &&+
        \left|g^{\text{FF}}_{\parallel}\right|^2
        \left| \frac{\alpha^{\text{tip}}_{zz}}{4\pi\varepsilon} \right|^2
        \frac{\pi^2}{z^2}
        \left[ \left|a_x(\xi,\varphi_\xi)\right|^2 + \left|a_y(\xi,\varphi_\xi)\right|^2 \right]
        \Bigg\},
    \end{eqnarray}
    \begin{eqnarray}\label{eq:IE2gx_step1}
        \mc{I}^{(E_{2g},x)}
        =
        |\alphabar^{(E_{2g})}|^2 |E_z (\omega)|^2
        &\Bigg\{&
        \left|g^{\text{FF}}_{z}\right|^2
        \left|
            \frac{\alpha^{\text{tip}}_{zz}}
            {4\pi \varepsilon}
        \right|^4
        \frac{\pi^2}{z^8}
        \left| h_{xy}(\xi,\varphi_\xi) + h_{yx}(\xi,\varphi_\xi) \right|^2
        \notag\\
        &&+
        \left|g^{\text{FF}}_{\parallel}\right|^2
        \left| \frac{\alpha^{\text{tip}}_{zz}}{4\pi\varepsilon} \right|^2
        \frac{\pi^2}{z^2}
        \left[ \left|a_x(\xi,\varphi_\xi)\right|^2 + \left|a_y(\xi,\varphi_\xi)\right|^2 \right]
        \Bigg\},
    \end{eqnarray}
    \begin{eqnarray}\label{eq:IE2gy_step1}
        \mc{I}^{(E_{2g},y)}
        =
        |\alphabar^{(E_{2g})}|^2 |E_z (\omega)|^2
        &\Bigg\{&
        \left|g^{\text{FF}}_{z}\right|^2
        \left|
            \frac{\alpha^{\text{tip}}_{zz}}
            {4\pi \varepsilon}
        \right|^4
        \frac{\pi^2}{z^8}
        \left| h_{xx}(\xi,\varphi_\xi) - h_{yy}(\xi,\varphi_\xi) \right|^2
        \notag\\
        &&+
        \left|g^{\text{FF}}_{\parallel}\right|^2
        \left| \frac{\alpha^{\text{tip}}_{zz}}{4\pi\varepsilon} \right|^2
        \frac{\pi^2}{z^2}
        \left[ \left|a_x(\xi,\varphi_\xi)\right|^2 + \left|a_y(\xi,\varphi_\xi)\right|^2 \right]
        \Bigg\}.
    \end{eqnarray}
\end{subequations}

Crucially, we have to work on the angular dependence of $\mc{I}^{(\eta)}$, that is, the dependence on variable $\varphi_\xi$.
The first simplification we do in this sense is that, using Eq.~(\ref{eq:a_analytical}), one can write
\begin{eqnarray}
    |a_x(\xi,\varphi_\xi)|^2
    + |a_y(\xi,\varphi_\xi)|^2
    &=&
    4
    \xi^2 e^{-2\xi}
    \left[ |\psi_x(\varphi_\xi)|^2 + |\psi_y(\varphi_\xi)|^2 \right]
    \notag\\
    &=&
    4
    \xi^2 e^{-2\xi}
    ,
\end{eqnarray}
meaning that the sample plane polarization in the ST contribution is independent of $\varphi_\xi$ for all modes.
We also note that, from Eq.~(\ref{eq:h_analytical}) and (\ref{eq:a_analytical}) all the $z$ contribution in Eq.~(\ref{eq:Ieta_step1}) is also independent of $\varphi_\xi$,
\begin{subequations}
    \begin{equation}
        h_{zz}(\xi, \varphi_\xi)
        =
        2
            \left[
                \frac{9}{384} \xi^4 K_4(\xi)
                - \frac{6}{48} \xi^3 K_3(\xi)
                + \frac{1}{8} \xi^2 K_2(\xi)
            \right],
    \end{equation}
    \begin{equation}
        a_z(\xi,\varphi_\xi)
        =
        2 \xi e^{-\xi}
        .
    \end{equation}
\end{subequations}
We are left with
\begin{subequations}
    \begin{equation}
        h_{xx}(\xi, \varphi_\xi)
        + h_{yy}(\xi, \varphi_\xi)
        =
        2 \frac{9}{384} \left[ 2 \xi^3 K_3(\xi) -\xi^4 K_2(\xi) \right] ,
    \end{equation}
    \begin{equation}
        h_{xy}(\xi, \varphi_\xi)
        + h_{yx}(\xi, \varphi_\xi)
        =
        - 2 \sin(2\varphi_\xi)
        \frac{9}{384} \xi^4 K_2(\xi) ,
    \end{equation}
    \begin{equation}
        h_{xx}(\xi, \varphi_\xi)
        - h_{yy}(\xi, \varphi_\xi)
        =
        - 2 \cos(2\varphi_\xi)
        \frac{9}{384} \xi^4 K_2(\xi) ,
    \end{equation}
\end{subequations}
where we used the definition of $\tau_{ij}(\varphi_\xi)$, Eq.~(\ref{eq:tau_def}).

Substitution of the above equations in Eq.~(\ref{eq:Ieta_step1}) yields
\begin{subequations}\label{eq:Ieta_step2}
    \begin{eqnarray}\label{eq:IA1_step2}
        \mc{I}^{(A_1)}
        =
        |\alphabar^{(A_1)}|^2 |E_z (\omega)|^2
        &\Bigg\{&
        \left|g^{\text{FF}}_{z}\right|^2
        \Bigg|
        \left(
            \frac{\alpha^{\text{tip}}_{zz}}
            {4\pi \varepsilon}
        \right)^2
        \frac{\pi}{z^4}
        \left[ \frac{3}{32} \xi^3 K_3(\xi) -\frac{3}{64} \xi^4 K_2(\xi) \right]
        \notag\\
        &&+ \nu \left(
            \frac{\alpha^{\text{tip}}_{zz}}
            {4\pi \varepsilon}
        \right)^2
        \frac{\pi}{z^4}
        \left[
            \frac{3}{64} \xi^4 K_4(\xi)
            - \frac{1}{4} \xi^3 K_3(\xi)
            + \frac{1}{4} \xi^2 K_2(\xi)
        \right]
        \notag\\
        &&+
        4 \nu
            \left( \frac{\alpha^{\text{tip}}_{zz}}{4\pi\varepsilon} \right)
            \frac{\pi}{z}
            \xi e^{-\xi}
        \Bigg|^2
        \notag\\
        &&+
        4 \left|g^{\text{FF}}_{\parallel}\right|^2
        \left| \frac{\alpha^{\text{tip}}_{zz}}{4\pi\varepsilon} \right|^2
        \frac{\pi^2}{z^2}
        \xi^2 e^{-2\xi}
        \Bigg\},
    \end{eqnarray}
    \begin{eqnarray}\label{eq:IE2gx_step2}
        \mc{I}^{(E_{2g},x)}
        =
        |\alphabar^{(E_{2g})}|^2 |E_z (\omega)|^2
        &\Bigg\{&
        4 \left|g^{\text{FF}}_{z}\right|^2
        \left|
            \frac{\alpha^{\text{tip}}_{zz}}
            {4\pi \varepsilon}
        \right|^4
        \frac{\pi^2}{z^8}
        \sin^2(2\varphi_\xi)
        \left(\frac{3}{128}\right)^2 \xi^8 [K_2(\xi)]^2
        \notag\\
        &&+
        4 \left|g^{\text{FF}}_{\parallel}\right|^2
        \left| \frac{\alpha^{\text{tip}}_{zz}}{4\pi\varepsilon} \right|^2
        \frac{\pi^2}{z^2}
        \xi^2 e^{-2\xi}
        \Bigg\},
    \end{eqnarray}
    \begin{eqnarray}\label{eq:IE2gy_step2}
        \mc{I}^{(E_{2g},y)}
        =
        |\alphabar^{(E_{2g})}|^2 |E_z (\omega)|^2
        &\Bigg\{&
        4 \left|g^{\text{FF}}_{z}\right|^2
        \left|
            \frac{\alpha^{\text{tip}}_{zz}}
            {4\pi \varepsilon}
        \right|^4
        \frac{\pi^2}{z^8}
        \cos^2(2\varphi_\xi)
        \left(\frac{3}{128}\right)^2 \xi^8 [K_2(\xi)]^2
        \notag\\
        &&+
        4 \left|g^{\text{FF}}_{\parallel}\right|^2
        \left| \frac{\alpha^{\text{tip}}_{zz}}{4\pi\varepsilon} \right|^2
        \frac{\pi^2}{z^2}
        \xi^2 e^{-2\xi}
        \Bigg\}.
    \end{eqnarray}
\end{subequations}

Having calculated $\mc{I}^{(\eta)}$, we can now calculate $\mc{J}^{(\eta)} (\xi;z) \equiv \int_0^{2\pi} \mc{I}^{(\eta)}(\xi,\varphi_\xi; z) d\varphi_\xi$,
which evaluates to
\begin{subequations}\label{eq:app:Jeta_results}
    \begin{eqnarray}
        \mc{J}^{(A_1)}(\xi;z)
        =
        2\pi |\alphabar^{(A_1)}|^2 |E_z (\omega)|^2
        &\Bigg\{&
        \left|g^{\text{FF}}_{z}\right|^2
        \Bigg|
        \left(
            \frac{\alpha^{\text{tip}}_{zz}}
            {4\pi \varepsilon}
        \right)^2
        \frac{2\pi}{z^4}
        \left( \frac{9}{384} \right)
        \left[  2 \xi^3 K_3(\xi) - \xi^4 K_2(\xi) \right]
        \notag\\
        &&+ \nu \left(
            \frac{\alpha^{\text{tip}}_{zz}}
            {4\pi \varepsilon}
        \right)^2
        \frac{2\pi}{z^4}
        \left[
            \frac{9}{384} \xi^4 K_4(\xi)
            - \frac{6}{48} \xi^3 K_3(\xi)
            + \frac{1}{8} \xi^2 K_2(\xi)
        \right]
        \notag\\
        &&+
        2 \nu
            \left( \frac{\alpha^{\text{tip}}_{zz}}{4\pi\varepsilon} \right)
            \frac{2 \pi}{z}
            \xi e^{-\xi}
        \Bigg|^2
        \notag\\
        &&+
        \left|g^{\text{FF}}_{\parallel}\right|^2
        \left|
            \left( \frac{\alpha^{\text{tip}}_{zz}}{4\pi\varepsilon} \right)
            \frac{2\pi}{z}
            \xi e^{-\xi}
        \right|^2
        \Bigg\},
    \end{eqnarray}
    \begin{eqnarray}
        \mc{J}^{(E_{2g},x)}(\xi;z)
        =
        \mc{J}^{(E_{2g},y)}(\xi;z)&&
        \notag\\
        =
        2\pi |\alphabar^{(E_{2g})}|^2 |E_z (\omega)|^2
        &\Bigg\{&
        \frac{1}{2} \left|g^{\text{FF}}_{z}\right|^2
        \left|
            \left( \frac{\alpha^{\text{tip}}_{zz}}{4\pi \varepsilon} \right)^2
            \frac{2 \pi}{z^4}
            \left(\frac{3}{128}\right) \xi^4 K_2(\xi)
        \right|^2
        \notag\\
        &&+
        \left|g^{\text{FF}}_{\parallel}\right|^2
        \left|
            \left( \frac{\alpha^{\text{tip}}_{zz}}{4\pi \varepsilon} \right)
            \frac{2\pi}{z}
            \xi e^{-\xi}
        \right|^2
        \Bigg\}.
    \end{eqnarray}
\end{subequations}


\begin{thebibliography}{29}%
\makeatletter
\providecommand \@ifxundefined [1]{%
 \@ifx{#1\undefined}
}%
\providecommand \@ifnum [1]{%
 \ifnum #1\expandafter \@firstoftwo
 \else \expandafter \@secondoftwo
 \fi
}%
\providecommand \@ifx [1]{%
 \ifx #1\expandafter \@firstoftwo
 \else \expandafter \@secondoftwo
 \fi
}%
\providecommand \natexlab [1]{#1}%
\providecommand \enquote  [1]{``#1''}%
\providecommand \bibnamefont  [1]{#1}%
\providecommand \bibfnamefont [1]{#1}%
\providecommand \citenamefont [1]{#1}%
\providecommand \href@noop [0]{\@secondoftwo}%
\providecommand \href [0]{\begingroup \@sanitize@url \@href}%
\providecommand \@href[1]{\@@startlink{#1}\@@href}%
\providecommand \@@href[1]{\endgroup#1\@@endlink}%
\providecommand \@sanitize@url [0]{\catcode `\\12\catcode `\$12\catcode `\&12\catcode `\#12\catcode `\^12\catcode `\_12\catcode `\%12\relax}%
\providecommand \@@startlink[1]{}%
\providecommand \@@endlink[0]{}%
\providecommand \url  [0]{\begingroup\@sanitize@url \@url }%
\providecommand \@url [1]{\endgroup\@href {#1}{\urlprefix }}%
\providecommand \urlprefix  [0]{URL }%
\providecommand \Eprint [0]{\href }%
\providecommand \doibase [0]{https://doi.org/}%
\providecommand \selectlanguage [0]{\@gobble}%
\providecommand \bibinfo  [0]{\@secondoftwo}%
\providecommand \bibfield  [0]{\@secondoftwo}%
\providecommand \translation [1]{[#1]}%
\providecommand \BibitemOpen [0]{}%
\providecommand \bibitemStop [0]{}%
\providecommand \bibitemNoStop [0]{.\EOS\space}%
\providecommand \EOS [0]{\spacefactor3000\relax}%
\providecommand \BibitemShut  [1]{\csname bibitem#1\endcsname}%
\let\auto@bib@innerbib\@empty
\bibitem [{\citenamefont {Cardona}(1983)}]{cardona1}%
  \BibitemOpen
  \bibinfo {editor} {\bibfnamefont {M.}~\bibnamefont {Cardona}},\ ed.,\ \href@noop {} {\emph {\bibinfo {title} {Light scattering in solids {I}}}}\ (\bibinfo  {publisher} {Springer-Verlag},\ \bibinfo {address} {Berlin},\ \bibinfo {year} {1983})\BibitemShut {NoStop}%
\bibitem [{\citenamefont {Bloembergen}(1996)}]{bloembergen}%
  \BibitemOpen
  \bibfield  {author} {\bibinfo {author} {\bibfnamefont {N.}~\bibnamefont {Bloembergen}},\ }\href@noop {} {\emph {\bibinfo {title} {Nonlinear optics}}},\ \bibinfo {edition} {4th}\ ed.\ (\bibinfo  {publisher} {World Scientific},\ \bibinfo {address} {New York},\ \bibinfo {year} {1996})\BibitemShut {NoStop}%
\bibitem [{\citenamefont {Novotny}\ and\ \citenamefont {Hecht}(2012)}]{novotny}%
  \BibitemOpen
  \bibfield  {author} {\bibinfo {author} {\bibfnamefont {L.}~\bibnamefont {Novotny}}\ and\ \bibinfo {author} {\bibfnamefont {B.}~\bibnamefont {Hecht}},\ }\href@noop {} {\emph {\bibinfo {title} {Principles of nano-optics}}},\ \bibinfo {edition} {2nd}\ ed.\ (\bibinfo  {publisher} {Cambridge University Press},\ \bibinfo {address} {Cambridge},\ \bibinfo {year} {2012})\BibitemShut {NoStop}%
\bibitem [{\citenamefont {Kawata}\ and\ \citenamefont {Shalaev}(2007)}]{kawata_shalaev}%
  \BibitemOpen
  \bibinfo {editor} {\bibfnamefont {S.}~\bibnamefont {Kawata}}\ and\ \bibinfo {editor} {\bibfnamefont {V.~M.}\ \bibnamefont {Shalaev}},\ eds.,\ \href@noop {} {\emph {\bibinfo {title} {Tip enhancement}}}\ (\bibinfo  {publisher} {Elsevier},\ \bibinfo {address} {Amsterdam},\ \bibinfo {year} {2007})\BibitemShut {NoStop}%
\bibitem [{\citenamefont {Jorio}\ \emph {et~al.}(2024)\citenamefont {Jorio}, \citenamefont {Nadas}, \citenamefont {Pereira}, \citenamefont {Rabelo}, \citenamefont {Gadelha}, \citenamefont {Vasconcelos}, \citenamefont {Zhang}, \citenamefont {Miyata}, \citenamefont {Saito}, \citenamefont {Costa},\ and\ \citenamefont {Cançado}}]{jorio2024}%
  \BibitemOpen
  \bibfield  {author} {\bibinfo {author} {\bibfnamefont {A.}~\bibnamefont {Jorio}}, \bibinfo {author} {\bibfnamefont {R.}~\bibnamefont {Nadas}}, \bibinfo {author} {\bibfnamefont {A.~G.}\ \bibnamefont {Pereira}}, \bibinfo {author} {\bibfnamefont {C.}~\bibnamefont {Rabelo}}, \bibinfo {author} {\bibfnamefont {A.~C.}\ \bibnamefont {Gadelha}}, \bibinfo {author} {\bibfnamefont {T.~L.}\ \bibnamefont {Vasconcelos}}, \bibinfo {author} {\bibfnamefont {W.}~\bibnamefont {Zhang}}, \bibinfo {author} {\bibfnamefont {Y.}~\bibnamefont {Miyata}}, \bibinfo {author} {\bibfnamefont {R.}~\bibnamefont {Saito}}, \bibinfo {author} {\bibfnamefont {M.~D.~D.}\ \bibnamefont {Costa}},\ and\ \bibinfo {author} {\bibfnamefont {L.~G.}\ \bibnamefont {Cançado}},\ }\bibfield  {title} {\bibinfo {title} {Nano-{R}aman spectroscopy of 2{D} materials},\ }\href {https://doi.org/10.1088/2053-1583/ad42ad} {\bibfield  {journal} {\bibinfo  {journal} {2D Mater.}\ }\textbf {\bibinfo {volume} {11}},\ \bibinfo {pages} {033003} (\bibinfo {year} {2024})}\BibitemShut {NoStop}%
\bibitem [{\citenamefont {Bao}\ \emph {et~al.}(2024)\citenamefont {Bao}, \citenamefont {Zhu}, \citenamefont {Zhao}, \citenamefont {Chen}, \citenamefont {Wang},\ and\ \citenamefont {Ren}}]{bao2024}%
  \BibitemOpen
  \bibfield  {author} {\bibinfo {author} {\bibfnamefont {Y.-F.}\ \bibnamefont {Bao}}, \bibinfo {author} {\bibfnamefont {M.-Y.}\ \bibnamefont {Zhu}}, \bibinfo {author} {\bibfnamefont {X.-J.}\ \bibnamefont {Zhao}}, \bibinfo {author} {\bibfnamefont {H.-X.}\ \bibnamefont {Chen}}, \bibinfo {author} {\bibfnamefont {X.~W.}\ \bibnamefont {Wang}},\ and\ \bibinfo {author} {\bibfnamefont {B.}~\bibnamefont {Ren}},\ }\bibfield  {title} {\bibinfo {title} {Nanoscale chemical characterization of materials and interfaces by tip-enhanced {R}aman spectroscopy},\ }\bibfield  {journal} {\bibinfo  {journal} {Chem. Soc. Rev.}\ }\href {https://doi.org/10.1039/D4CS00588K} {10.1039/D4CS00588K} (\bibinfo {year} {2024})\BibitemShut {NoStop}%
\bibitem [{\citenamefont {Beams}\ \emph {et~al.}(2014)\citenamefont {Beams}, \citenamefont {Can\ifmmode~\mbox{\c{c}}\else \c{c}\fi{}ado}, \citenamefont {Oh}, \citenamefont {Jorio},\ and\ \citenamefont {Novotny}}]{beams2014}%
  \BibitemOpen
  \bibfield  {author} {\bibinfo {author} {\bibfnamefont {R.}~\bibnamefont {Beams}}, \bibinfo {author} {\bibfnamefont {L.~G.}\ \bibnamefont {Can\ifmmode~\mbox{\c{c}}\else \c{c}\fi{}ado}}, \bibinfo {author} {\bibfnamefont {S.-H.}\ \bibnamefont {Oh}}, \bibinfo {author} {\bibfnamefont {A.}~\bibnamefont {Jorio}},\ and\ \bibinfo {author} {\bibfnamefont {L.}~\bibnamefont {Novotny}},\ }\bibfield  {title} {\bibinfo {title} {Spatial coherence in near-field {R}aman scattering},\ }\href {https://doi.org/10.1103/PhysRevLett.113.186101} {\bibfield  {journal} {\bibinfo  {journal} {Phys. Rev. Lett.}\ }\textbf {\bibinfo {volume} {113}},\ \bibinfo {pages} {186101} (\bibinfo {year} {2014})}\BibitemShut {NoStop}%
\bibitem [{\citenamefont {Nadas}\ \emph {et~al.}(2025)\citenamefont {Nadas}, \citenamefont {Corr\^ea}, \citenamefont {Cançado},\ and\ \citenamefont {Jorio}}]{nadas2025}%
  \BibitemOpen
  \bibfield  {author} {\bibinfo {author} {\bibfnamefont {R.}~\bibnamefont {Nadas}}, \bibinfo {author} {\bibfnamefont {R.}~\bibnamefont {Corr\^ea}}, \bibinfo {author} {\bibfnamefont {L.~G.}\ \bibnamefont {Cançado}},\ and\ \bibinfo {author} {\bibfnamefont {A.}~\bibnamefont {Jorio}},\ }\bibfield  {title} {\bibinfo {title} {Tip-enhanced {R}aman spectroscopy coherence length of 2{D} materials: An application to graphene},\ }\href {https://doi.org/https://doi.org/10.1002/pssb.202400287} {\bibfield  {journal} {\bibinfo  {journal} {Phys. Status Solidi B}\ }\textbf {\bibinfo {volume} {262}},\ \bibinfo {pages} {2400287} (\bibinfo {year} {2025})}\BibitemShut {NoStop}%
\bibitem [{\citenamefont {Alencar}\ \emph {et~al.}(2019)\citenamefont {Alencar}, \citenamefont {Rabelo}, \citenamefont {Miranda}, \citenamefont {Vasconcelos}, \citenamefont {Oliveira}, \citenamefont {Ribeiro}, \citenamefont {Públio}, \citenamefont {Ribeiro-Soares}, \citenamefont {Filho}, \citenamefont {Can{\c{c}}ado},\ and\ \citenamefont {Jorio}}]{alencar2019}%
  \BibitemOpen
  \bibfield  {author} {\bibinfo {author} {\bibfnamefont {R.~S.}\ \bibnamefont {Alencar}}, \bibinfo {author} {\bibfnamefont {C.}~\bibnamefont {Rabelo}}, \bibinfo {author} {\bibfnamefont {H.~L.~S.}\ \bibnamefont {Miranda}}, \bibinfo {author} {\bibfnamefont {T.~L.}\ \bibnamefont {Vasconcelos}}, \bibinfo {author} {\bibfnamefont {B.~S.}\ \bibnamefont {Oliveira}}, \bibinfo {author} {\bibfnamefont {A.}~\bibnamefont {Ribeiro}}, \bibinfo {author} {\bibfnamefont {B.~C.}\ \bibnamefont {Públio}}, \bibinfo {author} {\bibfnamefont {J.}~\bibnamefont {Ribeiro-Soares}}, \bibinfo {author} {\bibfnamefont {A.~G.~S.}\ \bibnamefont {Filho}}, \bibinfo {author} {\bibfnamefont {L.~G.}\ \bibnamefont {Can{\c{c}}ado}},\ and\ \bibinfo {author} {\bibfnamefont {A.}~\bibnamefont {Jorio}},\ }\bibfield  {title} {\bibinfo {title} {Probing spatial phonon correlation length in post-transition metal monochalcogenide {G}a{S} using tip-enhanced {R}aman spectroscopy},\ }\href {https://doi.org/10.1021/acs.nanolett.9b02974} {\bibfield  {journal} {\bibinfo  {journal} {Nano Lett.}\ }\textbf {\bibinfo {volume} {19}},\ \bibinfo {pages} {7357} (\bibinfo {year} {2019})}\BibitemShut {NoStop}%
\bibitem [{\citenamefont {Can\ifmmode~\mbox{\c{c}}\else \c{c}\fi{}ado}\ \emph {et~al.}(2014)\citenamefont {Can\ifmmode~\mbox{\c{c}}\else \c{c}\fi{}ado}, \citenamefont {Beams}, \citenamefont {Jorio},\ and\ \citenamefont {Novotny}}]{cancado2014}%
  \BibitemOpen
  \bibfield  {author} {\bibinfo {author} {\bibfnamefont {L.~G.}\ \bibnamefont {Can\ifmmode~\mbox{\c{c}}\else \c{c}\fi{}ado}}, \bibinfo {author} {\bibfnamefont {R.}~\bibnamefont {Beams}}, \bibinfo {author} {\bibfnamefont {A.}~\bibnamefont {Jorio}},\ and\ \bibinfo {author} {\bibfnamefont {L.}~\bibnamefont {Novotny}},\ }\bibfield  {title} {\bibinfo {title} {Theory of spatial coherence in near-field {R}aman scattering},\ }\href {https://doi.org/10.1103/PhysRevX.4.031054} {\bibfield  {journal} {\bibinfo  {journal} {Phys. Rev. X}\ }\textbf {\bibinfo {volume} {4}},\ \bibinfo {pages} {031054} (\bibinfo {year} {2014})}\BibitemShut {NoStop}%
\bibitem [{\citenamefont {Publio}\ \emph {et~al.}(2022)\citenamefont {Publio}, \citenamefont {Oliveira}, \citenamefont {Rabelo}, \citenamefont {Miranda}, \citenamefont {Vasconcelos}, \citenamefont {Jorio},\ and\ \citenamefont {Can\ifmmode~\mbox{\c{c}}\else \c{c}\fi{}ado}}]{publio2022}%
  \BibitemOpen
  \bibfield  {author} {\bibinfo {author} {\bibfnamefont {B.~C.}\ \bibnamefont {Publio}}, \bibinfo {author} {\bibfnamefont {B.~S.}\ \bibnamefont {Oliveira}}, \bibinfo {author} {\bibfnamefont {C.}~\bibnamefont {Rabelo}}, \bibinfo {author} {\bibfnamefont {H.}~\bibnamefont {Miranda}}, \bibinfo {author} {\bibfnamefont {T.~L.}\ \bibnamefont {Vasconcelos}}, \bibinfo {author} {\bibfnamefont {A.}~\bibnamefont {Jorio}},\ and\ \bibinfo {author} {\bibfnamefont {L.~G.}\ \bibnamefont {Can\ifmmode~\mbox{\c{c}}\else \c{c}\fi{}ado}},\ }\bibfield  {title} {\bibinfo {title} {Inclusion of the sample-tip interaction term in the theory of tip-enhanced {R}aman spectroscopy},\ }\href {https://doi.org/10.1103/PhysRevB.105.235414} {\bibfield  {journal} {\bibinfo  {journal} {Phys. Rev. B}\ }\textbf {\bibinfo {volume} {105}},\ \bibinfo {pages} {235414} (\bibinfo {year} {2022})}\BibitemShut {NoStop}%
\bibitem [{\citenamefont {Jin}\ \emph {et~al.}(2020)\citenamefont {Jin}, \citenamefont {Zheng}, \citenamefont {Ding}, \citenamefont {Zhu}, \citenamefont {Wang},\ and\ \citenamefont {Huang}}]{raman_tensor_mose2}%
  \BibitemOpen
  \bibfield  {author} {\bibinfo {author} {\bibfnamefont {M.}~\bibnamefont {Jin}}, \bibinfo {author} {\bibfnamefont {W.}~\bibnamefont {Zheng}}, \bibinfo {author} {\bibfnamefont {Y.}~\bibnamefont {Ding}}, \bibinfo {author} {\bibfnamefont {Y.}~\bibnamefont {Zhu}}, \bibinfo {author} {\bibfnamefont {W.}~\bibnamefont {Wang}},\ and\ \bibinfo {author} {\bibfnamefont {F.}~\bibnamefont {Huang}},\ }\bibfield  {title} {\bibinfo {title} {Raman tensor of van der {W}aals {M}o{S}e\textsubscript{2}},\ }\href {https://doi.org/10.1021/acs.jpclett.0c01183} {\bibfield  {journal} {\bibinfo  {journal} {J. Phys. Chem. Lett.}\ }\textbf {\bibinfo {volume} {11}},\ \bibinfo {pages} {4311} (\bibinfo {year} {2020})}\BibitemShut {NoStop}%
\bibitem [{\citenamefont {Pratama}\ \emph {et~al.}(2019)\citenamefont {Pratama}, \citenamefont {Ukhtary},\ and\ \citenamefont {Saito}}]{pratama2019}%
  \BibitemOpen
  \bibfield  {author} {\bibinfo {author} {\bibfnamefont {F.~R.}\ \bibnamefont {Pratama}}, \bibinfo {author} {\bibfnamefont {M.~S.}\ \bibnamefont {Ukhtary}},\ and\ \bibinfo {author} {\bibfnamefont {R.}~\bibnamefont {Saito}},\ }\bibfield  {title} {\bibinfo {title} {Non-vertical optical transition in near-field enhanced spectroscopy of graphene},\ }\href {https://doi.org/10.1088/1361-648X/ab1335} {\bibfield  {journal} {\bibinfo  {journal} {J. Phys.: Condens. Matter}\ }\textbf {\bibinfo {volume} {31}},\ \bibinfo {pages} {265701} (\bibinfo {year} {2019})}\BibitemShut {NoStop}%
\bibitem [{Note1()}]{Note1}%
  \BibitemOpen
  \bibinfo {note} {We do not use repeated index summation convention because it may get ambiguous later.}\BibitemShut {Stop}%
\bibitem [{\citenamefont {Mandel}\ and\ \citenamefont {Wolf}(1995)}]{mandelwolf}%
  \BibitemOpen
  \bibfield  {author} {\bibinfo {author} {\bibfnamefont {L.}~\bibnamefont {Mandel}}\ and\ \bibinfo {author} {\bibfnamefont {E.}~\bibnamefont {Wolf}},\ }\href@noop {} {\emph {\bibinfo {title} {Optical coherence and quantum optics}}}\ (\bibinfo  {publisher} {Cambridge University Press},\ {New York},\ \bibinfo {year} {1995})\BibitemShut {NoStop}%
\bibitem [{\citenamefont {Tan}(2019)}]{tan_raman}%
  \BibitemOpen
  \bibinfo {editor} {\bibfnamefont {P.-H.}\ \bibnamefont {Tan}},\ ed.,\ \href@noop {} {\emph {\bibinfo {title} {Raman spectroscopy of Two-dimensional materials}}}\ (\bibinfo  {publisher} {Springer Nature},\ \bibinfo {address} {Singapore},\ \bibinfo {year} {2019})\BibitemShut {NoStop}%
\bibitem [{\citenamefont {Park}\ \emph {et~al.}(2016)\citenamefont {Park}, \citenamefont {Khatib}, \citenamefont {Kravtsov}, \citenamefont {Clark}, \citenamefont {Xu},\ and\ \citenamefont {Raschke}}]{park2016}%
  \BibitemOpen
  \bibfield  {author} {\bibinfo {author} {\bibfnamefont {K.-D.}\ \bibnamefont {Park}}, \bibinfo {author} {\bibfnamefont {O.}~\bibnamefont {Khatib}}, \bibinfo {author} {\bibfnamefont {V.}~\bibnamefont {Kravtsov}}, \bibinfo {author} {\bibfnamefont {G.}~\bibnamefont {Clark}}, \bibinfo {author} {\bibfnamefont {X.}~\bibnamefont {Xu}},\ and\ \bibinfo {author} {\bibfnamefont {M.~B.}\ \bibnamefont {Raschke}},\ }\bibfield  {title} {\bibinfo {title} {Hybrid tip-enhanced nanospectroscopy and nanoimaging of monolayer {WSe2} with local strain control},\ }\href {https://doi.org/10.1021/acs.nanolett.6b00238} {\bibfield  {journal} {\bibinfo  {journal} {Nano Lett.}\ }\textbf {\bibinfo {volume} {16}},\ \bibinfo {pages} {2621} (\bibinfo {year} {2016})}\BibitemShut {NoStop}%
\bibitem [{\citenamefont {Jorio}\ \emph {et~al.}(2012)\citenamefont {Jorio}, \citenamefont {Saito}, \citenamefont {Dresselhaus},\ and\ \citenamefont {Dresselhaus}}]{jorio_graphene}%
  \BibitemOpen
  \bibfield  {author} {\bibinfo {author} {\bibfnamefont {A.}~\bibnamefont {Jorio}}, \bibinfo {author} {\bibfnamefont {R.}~\bibnamefont {Saito}}, \bibinfo {author} {\bibfnamefont {G.}~\bibnamefont {Dresselhaus}},\ and\ \bibinfo {author} {\bibfnamefont {M.~S.}\ \bibnamefont {Dresselhaus}},\ }\href@noop {} {\emph {\bibinfo {title} {Raman Spectroscopy in Graphene Related Systems}}}\ (\bibinfo  {publisher} {Wiley-VCH Verlag},\ \bibinfo {address} {Weinheim},\ \bibinfo {year} {2012})\BibitemShut {NoStop}%
\bibitem [{\citenamefont {Zhao}\ \emph {et~al.}(2013)\citenamefont {Zhao}, \citenamefont {Luo}, \citenamefont {Li}, \citenamefont {Zhang}, \citenamefont {Araujo}, \citenamefont {Gan}, \citenamefont {Wu}, \citenamefont {Zhang}, \citenamefont {Quek}, \citenamefont {Dresselhaus},\ and\ \citenamefont {Xiong}}]{zhao2013}%
  \BibitemOpen
  \bibfield  {author} {\bibinfo {author} {\bibfnamefont {Y.}~\bibnamefont {Zhao}}, \bibinfo {author} {\bibfnamefont {X.}~\bibnamefont {Luo}}, \bibinfo {author} {\bibfnamefont {H.}~\bibnamefont {Li}}, \bibinfo {author} {\bibfnamefont {J.}~\bibnamefont {Zhang}}, \bibinfo {author} {\bibfnamefont {P.~T.}\ \bibnamefont {Araujo}}, \bibinfo {author} {\bibfnamefont {C.~K.}\ \bibnamefont {Gan}}, \bibinfo {author} {\bibfnamefont {J.}~\bibnamefont {Wu}}, \bibinfo {author} {\bibfnamefont {H.}~\bibnamefont {Zhang}}, \bibinfo {author} {\bibfnamefont {S.~Y.}\ \bibnamefont {Quek}}, \bibinfo {author} {\bibfnamefont {M.~S.}\ \bibnamefont {Dresselhaus}},\ and\ \bibinfo {author} {\bibfnamefont {Q.}~\bibnamefont {Xiong}},\ }\bibfield  {title} {\bibinfo {title} {Interlayer breathing and shear modes in few-trilayer {M}o{S}2 and {WS}e2},\ }\href {https://doi.org/10.1021/nl304169w} {\bibfield  {journal} {\bibinfo  {journal} {Nano Lett.}\ }\textbf {\bibinfo {volume} {13}},\ \bibinfo {pages} {1007} (\bibinfo {year} {2013})}\BibitemShut {NoStop}%
\bibitem [{\citenamefont {Piessens}(2000)}]{piessens_hankel}%
  \BibitemOpen
  \bibfield  {author} {\bibinfo {author} {\bibfnamefont {R.}~\bibnamefont {Piessens}},\ }\bibfield  {title} {\bibinfo {title} {The {H}ankel transform},\ }in\ \href@noop {} {\emph {\bibinfo {booktitle} {The Transforms and Applications Handbook}}},\ \bibinfo {editor} {edited by\ \bibinfo {editor} {\bibfnamefont {A.~D.}\ \bibnamefont {Poularikas}}}\ (\bibinfo  {publisher} {CRC Press LLC},\ \bibinfo {address} {Boca Raton},\ \bibinfo {year} {2000})\ \bibinfo {edition} {2nd}\ ed.\BibitemShut {Stop}%
\bibitem [{\citenamefont {Miranda}\ \emph {et~al.}(2022)\citenamefont {Miranda}, \citenamefont {Monken}, \citenamefont {Campos}, \citenamefont {Vasconcelos}, \citenamefont {Rabelo}, \citenamefont {Archanjo}, \citenamefont {Almeida}, \citenamefont {Grieger}, \citenamefont {Backes}, \citenamefont {Jorio},\ and\ \citenamefont {Gustavo~Can{\c{c}}ado}}]{miranda2023}%
  \BibitemOpen
  \bibfield  {author} {\bibinfo {author} {\bibfnamefont {H.}~\bibnamefont {Miranda}}, \bibinfo {author} {\bibfnamefont {V.}~\bibnamefont {Monken}}, \bibinfo {author} {\bibfnamefont {J.~L.~E.}\ \bibnamefont {Campos}}, \bibinfo {author} {\bibfnamefont {T.~L.}\ \bibnamefont {Vasconcelos}}, \bibinfo {author} {\bibfnamefont {C.}~\bibnamefont {Rabelo}}, \bibinfo {author} {\bibfnamefont {B.~S.}\ \bibnamefont {Archanjo}}, \bibinfo {author} {\bibfnamefont {C.~M.}\ \bibnamefont {Almeida}}, \bibinfo {author} {\bibfnamefont {S.}~\bibnamefont {Grieger}}, \bibinfo {author} {\bibfnamefont {C.}~\bibnamefont {Backes}}, \bibinfo {author} {\bibfnamefont {A.}~\bibnamefont {Jorio}},\ and\ \bibinfo {author} {\bibfnamefont {L.}~\bibnamefont {Gustavo~Can{\c{c}}ado}},\ }\bibfield  {title} {\bibinfo {title} {Establishing the excitation field in tip-enhanced {R}aman spectroscopy to study nanostructures within two-dimensional systems},\ }\href {https://doi.org/10.1088/2053-1583/ac988f} {\bibfield  {journal} {\bibinfo  {journal} {2D Mater.}\ }\textbf {\bibinfo {volume} {10}},\ \bibinfo {pages} {015002} (\bibinfo {year} {2022})}\BibitemShut {NoStop}%
\bibitem [{\citenamefont {Can\ifmmode~\mbox{\c{c}}\else \c{c}\fi{}ado}\ \emph {et~al.}(2009)\citenamefont {Can\ifmmode~\mbox{\c{c}}\else \c{c}\fi{}ado}, \citenamefont {Jorio}, \citenamefont {Ismach}, \citenamefont {Joselevich}, \citenamefont {Hartschuh},\ and\ \citenamefont {Novotny}}]{cancado2009}%
  \BibitemOpen
  \bibfield  {author} {\bibinfo {author} {\bibfnamefont {L.~G.}\ \bibnamefont {Can\ifmmode~\mbox{\c{c}}\else \c{c}\fi{}ado}}, \bibinfo {author} {\bibfnamefont {A.}~\bibnamefont {Jorio}}, \bibinfo {author} {\bibfnamefont {A.}~\bibnamefont {Ismach}}, \bibinfo {author} {\bibfnamefont {E.}~\bibnamefont {Joselevich}}, \bibinfo {author} {\bibfnamefont {A.}~\bibnamefont {Hartschuh}},\ and\ \bibinfo {author} {\bibfnamefont {L.}~\bibnamefont {Novotny}},\ }\bibfield  {title} {\bibinfo {title} {Mechanism of near-field {R}aman enhancement in one-dimensional systems},\ }\href {https://doi.org/10.1103/PhysRevLett.103.186101} {\bibfield  {journal} {\bibinfo  {journal} {Phys. Rev. Lett.}\ }\textbf {\bibinfo {volume} {103}},\ \bibinfo {pages} {186101} (\bibinfo {year} {2009})}\BibitemShut {NoStop}%
\bibitem [{\citenamefont {Guimar\~aes}\ \emph {et~al.}(2025)\citenamefont {Guimar\~aes}, \citenamefont {Nadas}, \citenamefont {Alves}, \citenamefont {Zhang}, \citenamefont {Endo}, \citenamefont {Watanabe}, \citenamefont {Taniguchi}, \citenamefont {Saito}, \citenamefont {Miyata}, \citenamefont {Neves},\ and\ \citenamefont {Jorio}}]{guimaraes2025}%
  \BibitemOpen
  \bibfield  {author} {\bibinfo {author} {\bibfnamefont {J.~E.}\ \bibnamefont {Guimar\~aes}}, \bibinfo {author} {\bibfnamefont {R.}~\bibnamefont {Nadas}}, \bibinfo {author} {\bibfnamefont {R.}~\bibnamefont {Alves}}, \bibinfo {author} {\bibfnamefont {W.}~\bibnamefont {Zhang}}, \bibinfo {author} {\bibfnamefont {T.}~\bibnamefont {Endo}}, \bibinfo {author} {\bibfnamefont {K.}~\bibnamefont {Watanabe}}, \bibinfo {author} {\bibfnamefont {T.}~\bibnamefont {Taniguchi}}, \bibinfo {author} {\bibfnamefont {R.}~\bibnamefont {Saito}}, \bibinfo {author} {\bibfnamefont {Y.}~\bibnamefont {Miyata}}, \bibinfo {author} {\bibfnamefont {B.~R.~A.}\ \bibnamefont {Neves}},\ and\ \bibinfo {author} {\bibfnamefont {A.}~\bibnamefont {Jorio}},\ }\bibfield  {title} {\bibinfo {title} {Nano-{R}aman spectroscopy figure of merit and chemical analysis of contaminations in single-layer {M}o{S}e\textsubscript{2}},\ }\href {https://doi.org/10.1021/acsnano.5c08036} {\bibfield  {journal} {\bibinfo  {journal} {ACS Nano}\ }\textbf {\bibinfo {volume} {19}},\ \bibinfo {pages} {35438} (\bibinfo {year} {2025})}\BibitemShut {NoStop}%
\bibitem [{\citenamefont {Jackson}(1999)}]{jackson}%
  \BibitemOpen
  \bibfield  {author} {\bibinfo {author} {\bibfnamefont {J.~D.}\ \bibnamefont {Jackson}},\ }\href@noop {} {\emph {\bibinfo {title} {Classical Electrodynamics}}},\ \bibinfo {edition} {3rd}\ ed.\ (\bibinfo  {publisher} {John Wiley \& Sons},\ \bibinfo {address} {New York},\ \bibinfo {year} {1999})\BibitemShut {NoStop}%
\bibitem [{\citenamefont {Nadas}\ \emph {et~al.}(2023)\citenamefont {Nadas}, \citenamefont {Gadelha}, \citenamefont {Barbosa}, \citenamefont {Rabelo}, \citenamefont {de~Louren{\c{c}}o~e Vasconcelos}, \citenamefont {Monken}, \citenamefont {Portes}, \citenamefont {Watanabe}, \citenamefont {Taniguchi}, \citenamefont {Ramirez}, \citenamefont {Campos}, \citenamefont {Saito}, \citenamefont {Can{\c{c}}ado},\ and\ \citenamefont {Jorio}}]{nadas2023}%
  \BibitemOpen
  \bibfield  {author} {\bibinfo {author} {\bibfnamefont {R.~B.}\ \bibnamefont {Nadas}}, \bibinfo {author} {\bibfnamefont {A.~C.}\ \bibnamefont {Gadelha}}, \bibinfo {author} {\bibfnamefont {T.~C.}\ \bibnamefont {Barbosa}}, \bibinfo {author} {\bibfnamefont {C.}~\bibnamefont {Rabelo}}, \bibinfo {author} {\bibfnamefont {T.}~\bibnamefont {de~Louren{\c{c}}o~e Vasconcelos}}, \bibinfo {author} {\bibfnamefont {V.}~\bibnamefont {Monken}}, \bibinfo {author} {\bibfnamefont {A.~V.~R.}\ \bibnamefont {Portes}}, \bibinfo {author} {\bibfnamefont {K.}~\bibnamefont {Watanabe}}, \bibinfo {author} {\bibfnamefont {T.}~\bibnamefont {Taniguchi}}, \bibinfo {author} {\bibfnamefont {J.~C.}\ \bibnamefont {Ramirez}}, \bibinfo {author} {\bibfnamefont {L.~C.}\ \bibnamefont {Campos}}, \bibinfo {author} {\bibfnamefont {R.}~\bibnamefont {Saito}}, \bibinfo {author} {\bibfnamefont {L.~G.}\ \bibnamefont {Can{\c{c}}ado}},\ and\ \bibinfo {author} {\bibfnamefont {A.}~\bibnamefont {Jorio}},\ }\bibfield  {title} {\bibinfo {title} {Spatially coherent tip-enhanced raman spectroscopy measurements of electron–phonon interaction in a graphene device},\ }\href {https://doi.org/10.1021/acs.nanolett.3c00851} {\bibfield  {journal} {\bibinfo  {journal} {Nano Lett.}\ }\textbf {\bibinfo {volume} {23}},\ \bibinfo {pages} {8827} (\bibinfo {year} {2023})}\BibitemShut {NoStop}%
\bibitem [{\citenamefont {Neto}\ \emph {et~al.}(2019)\citenamefont {Neto}, \citenamefont {Rabelo}, \citenamefont {Cançado}, \citenamefont {Engel}, \citenamefont {Steiner},\ and\ \citenamefont {Jorio}}]{ribeironeto2019}%
  \BibitemOpen
  \bibfield  {author} {\bibinfo {author} {\bibfnamefont {A.~R.}\ \bibnamefont {Neto}}, \bibinfo {author} {\bibfnamefont {C.}~\bibnamefont {Rabelo}}, \bibinfo {author} {\bibfnamefont {L.~G.}\ \bibnamefont {Cançado}}, \bibinfo {author} {\bibfnamefont {M.}~\bibnamefont {Engel}}, \bibinfo {author} {\bibfnamefont {M.}~\bibnamefont {Steiner}},\ and\ \bibinfo {author} {\bibfnamefont {A.}~\bibnamefont {Jorio}},\ }\bibfield  {title} {\bibinfo {title} {Protocol and reference material for measuring the nanoantenna enhancement factor in tip-enhanced {R}aman spectroscopy},\ }in\ \href {https://doi.org/10.1109/INSCIT.2019.8868468} {\emph {\bibinfo {booktitle} {2019 4th International Symposium on Instrumentation Systems, Circuits and Transducers (INSCIT)}}}\ (\bibinfo {year} {2019})\ pp.\ \bibinfo {pages} {1--6}\BibitemShut {NoStop}%
\bibitem [{\citenamefont {Cuyt}\ \emph {et~al.}(2008)\citenamefont {Cuyt}, \citenamefont {Petersen}, \citenamefont {Verdonk}, \citenamefont {Waadeland},\ and\ \citenamefont {Jones}}]{cuyt}%
  \BibitemOpen
  \bibfield  {author} {\bibinfo {author} {\bibfnamefont {A.}~\bibnamefont {Cuyt}}, \bibinfo {author} {\bibfnamefont {V.~B.}\ \bibnamefont {Petersen}}, \bibinfo {author} {\bibfnamefont {B.}~\bibnamefont {Verdonk}}, \bibinfo {author} {\bibfnamefont {H.}~\bibnamefont {Waadeland}},\ and\ \bibinfo {author} {\bibfnamefont {W.~B.}\ \bibnamefont {Jones}},\ }\href@noop {} {\emph {\bibinfo {title} {Handbook of Continued Fractions for Special Functions}}}\ (\bibinfo  {publisher} {Springer},\ {Dordrecht},\ \bibinfo {year} {2008})\BibitemShut {NoStop}%
\bibitem [{\citenamefont {Davies}(2002)}]{davies_transforms}%
  \BibitemOpen
  \bibfield  {author} {\bibinfo {author} {\bibfnamefont {B.}~\bibnamefont {Davies}},\ }\href@noop {} {\emph {\bibinfo {title} {Integral transforms and their applications}}},\ \bibinfo {edition} {3rd}\ ed.\ (\bibinfo  {publisher} {Springer-Verlag},\ \bibinfo {address} {New York},\ \bibinfo {year} {2002})\BibitemShut {NoStop}%
\bibitem [{\citenamefont {Relton}(1965)}]{relton_bessel}%
  \BibitemOpen
  \bibfield  {author} {\bibinfo {author} {\bibfnamefont {F.~E.}\ \bibnamefont {Relton}},\ }\href@noop {} {\emph {\bibinfo {title} {Applied {B}essel functions}}}\ (\bibinfo  {publisher} {Dover Publications},\ \bibinfo {address} {New York},\ \bibinfo {year} {1965})\BibitemShut {NoStop}%
\end{thebibliography}
\end{document}